\documentclass[aps, superscriptaddress, twocolumn, pre]{revtex4-1}
\usepackage[font=small,labelfont=bf]{caption}
\usepackage[large]{subfigure }
\usepackage{amsmath,graphicx,color,epsfig,latexsym,bm,ulem,dutchcal,float,tikz,eucal, mathpazo,times, braket}
\usepackage[colorlinks=true, linkcolor = blue, urlcolor  = blue, citecolor = blue, anchorcolor = red]{hyperref}   

\usepackage{lipsum}
\usetikzlibrary{positioning}
\begin{document}
	\title{Heat current magnification in Classical and Quantum spin networks}
	\author{Vipul Upadhyay}\email{vipuupadhyay4@gmail.com}
	\affiliation{Department of Physics, Indian Institute of Technology Delhi, Hauz Khas 110 016, INDIA}
	\author{Poshika Gandhi}
	\affiliation{Institute of Physics, University of Freiburg, Hermann-Herder-Str. 3, 79104 Freiburg, Germany}
	\author{Rohit Juneja}
	\affiliation{Department of Physics, Indian Institute of Technology Delhi, Hauz Khas 110 016, INDIA}
	\author{Rahul Marathe}\email{maratherahul@physics.iitd.ac.in} \affiliation{Department of Physics, Indian Institute of Technology Delhi, Hauz Khas 110 016, INDIA}

	\begin{abstract}{We investigate heat current magnification due to asymmetry in the number of spins in two-branched classical and quantum spin systems. We begin by studying the classical Ising like spin models using Q2R and CCA dynamics and show that just the difference in the number of spins is not enough and some other source of asymmetry is required to observe heat current magnification. Unequal spin--spin interaction strength in the upper and lower branch is employed as a source of this asymmetry and it proves adequate for generating current magnification in both the models. Suitable physical motivation is then provided for current magnification in these systems, along with ways to control and manipulate magnification through various system parameters. We also study a five spin Quantum system with modified Heisenberg XXZ interaction and preserved magnetisation using the Redfield master equation. We show that it is possible to generate current magnification in this model by the asymmetry in the number of spins only. { Our results indicate that the onset of current magnification is accompanied by a dip in the total current  flowing through the system. On analysis it is revealed that this dip might occur because of the intersection of two non-degenerate energy levels for certain values of the asymmetry parameter in the modified  XXZ model. We deduce that the additional degeneracy and the ergodic constraint due to fixed magnetisation in the system 
				are the main reasons for current magnification and other atypical behaviors observed. We then use the concept of `ergotropy' to support these findings. Finally, for both the classical and quantum models, we see that current magnification is only observed when temperature gradient and intra-system interaction strength have similar order of energy.}}
	\end{abstract}
	\maketitle
	
	\section{Introduction} \label{sec::Introduction}
	
\begin{figure}[b]
		\centering
		\includegraphics[width=0.48\textwidth,height=0.1\textheight]{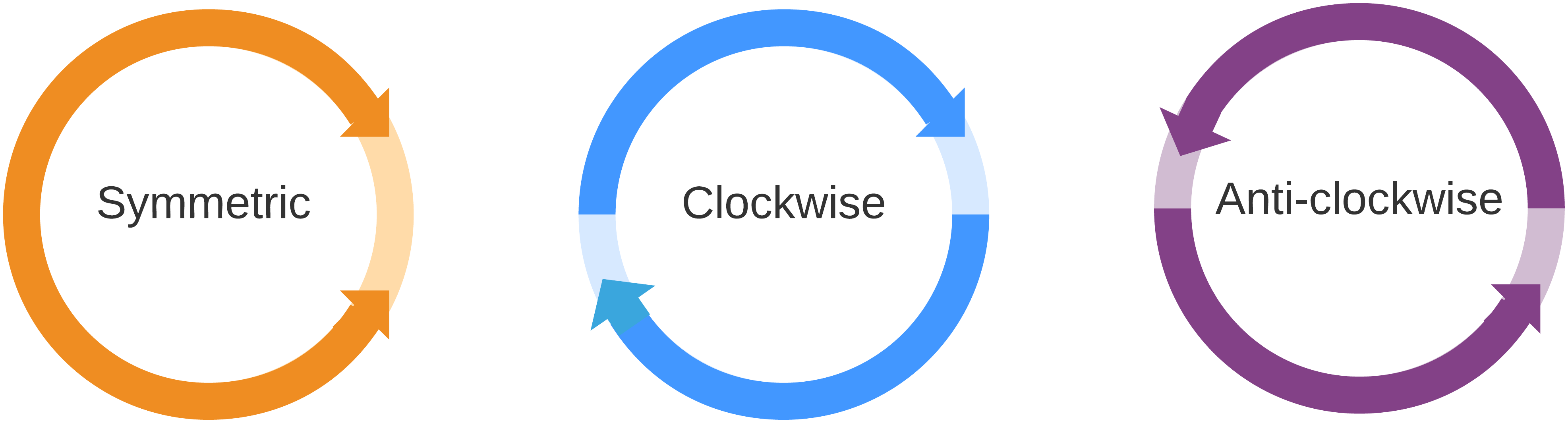}
		\caption{Possible ways for current to flow in a two-branch system (motivated from \cite{main_reference}) }
		\label{magnification}
	\end{figure}

	Heat currents can influence systems in a significant manner at the nano-scale and as a result, their control and manipulation at small scale is currently a subject of intense research in both the classical as well as quantum domain. These studies may be broadly divided into two categories, one dealing with finding ways to control heat currents by  modelling and manufacturing of small scale thermal devices like thermal diodes \cite{group_circulation,my,diode1,diode2,diode4,diode5,diode6,diode7,Zhang_2013,PhysRevLett.88.094302}, thermal transistors \cite{transistor1,transistor2,transistor3,transistor4,transistor5,transistor7,transistor8,transistor9,Nature460} etc., and the other more direct way of explicitly studying heat current transport and distribution in various systems, and developing transportation theories \cite{abhishek_dhar,arxiv1,arxiv2,ref2,ref3,ref4,ref6,ref7,ref8,e15062100,kirchberg2022}. Interestingly, for a multi branched system kept between two heat baths at different temperatures, it is observed that under certain conditions heat current in one of the branches may become larger than the total current flowing between the baths, thus leading to the phenomenon called the current magnification or circular current \cite{main_reference,geometry_circulation,Roy_2007,abraham_nitzan,rahul_sir,current_magnification2,current_magnification3,current_magnification4,current_magnification5,PhysRevLett.87.126801,zabey,Roy_2007}. Current magnification has been shown to exist in a wide variety of physical systems like spins \cite{main_reference}, molecules  \cite{abraham_nitzan,PhysRevLett.87.126801}, classical harmonic chains \cite{rahul_sir, zabey},  and metallic rings \cite{jayannavar2_fano,Roy_2007}. In general, it is observed that to get current magnification, the system must possess either rotational or reflection asymmetry \cite{main_reference}. For the simplest case of a system containing two branches, the heat current can flow in three possible ways, namely, parallel currents in the branches, clockwise circulating current, and anti--clockwise circulating current (see Fig. \ref{magnification}).
	In a recent study \cite{main_reference}, it was observed that it is possible to get circulating heat currents in Quantum spin systems with modified Heisenberg exchange interaction if the onsite magnetic field is inhomogeneous and the total magnetisation of the system is conserved. However, it is difficult to realise this system experimentally as applying different magnetic fields at different sites for such small systems is in general a difficult task.  Motivated by this and earlier studies on heat current magnification, here we study some simple classical and quantum models for heat current magnification in two-branched spin systems. In particular, we study the heat current flow in the classical Creutz Cellular automaton (CCA) and Q2R models described in \cite{Q2R,  cellular,CREUTZ198662,Ergodic_Schulte_1987,Q2R_STAUFFER2000113}. We consider systems with unequal branch spin numbers and with equal as well as unequal branch spin--spin interaction strengths. We find the required conditions for getting current magnification, and the range of parameters that optimise it. We also study a five spin Quantum system similar to \cite{main_reference} but with  spin number asymmetry in two branches. We use the Redfield master equation, and explore the heat flow and energetics in it. Similar to the classical models we find the parameter range  suitable for getting current magnification in this system. {The models that are studied in this manuscript can have possible experimental realizations for example in systems like quasi-1D Ising chains \cite{experiment_ising, wolf_2000} or quantum simulators like NMR \cite{Nature_experiment, NMR2}, quantum dots \cite{exp1, exp4}, trapped ions \cite{exp2, exp5}.}
	
	The paper is organized as follows. In Sec.~\ref{sec::Classical} we consider the classical spin models and provide analytical and numerical results. In Sec.~ \ref{sec::Quantum} we discuss the numerical \cite{faster_redfield} implementation of branch spin number asymmetry in the Quantum case by analysing our system by the Redfield master equation and highlight the relative ease for achieving current magnification in our model. We summarise our findings and conclude in Sec.~\ref{sec::Conclusion}.
	
	\section{Model and Classical Analysis} \label{sec::Classical}
	
	\begin{figure}[t]
		\centering
		\includegraphics[width=0.49\textwidth,height=0.15\textheight]{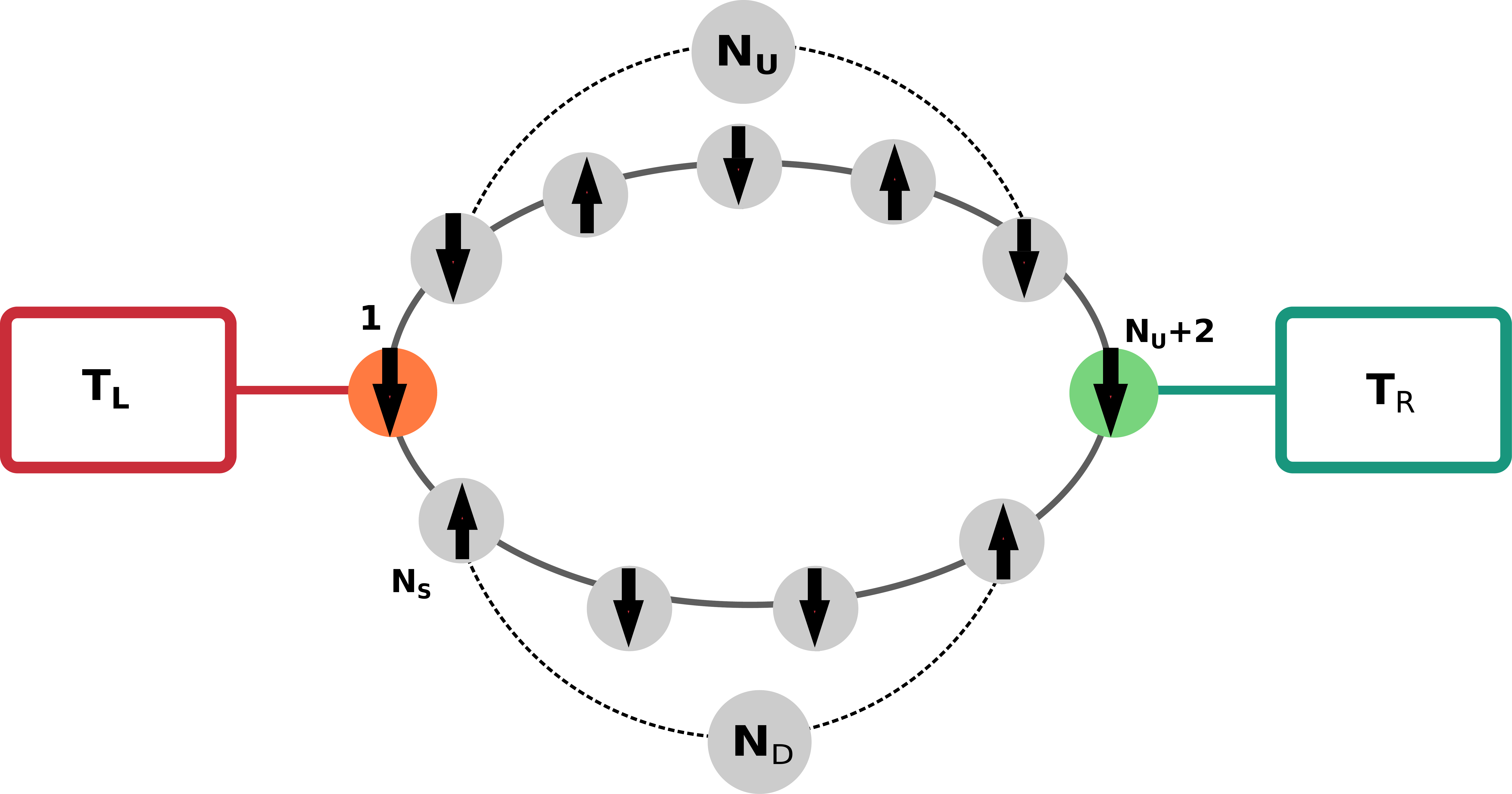}
		\caption{Schematic representation of the model system}
		\label{model1}
	\end{figure}
	We consider an Ising-like two-state spin chain with periodic boundary conditions as shown in Fig. \ref{model1}. This system is kept between two heat baths at temperatures $T_L$ and $T_R$ to its left and right respectively. The system interacts with the heat baths at two individual nodes such that only the nodal spins are in direct contact with them while the bulk spins are not. This results in the formation of two spin branches between the baths as shown in Fig. \ref{model1}. We use $N_U (N_D)$  for the number of spins in the upper (lower) branch which are not in contact with the heat baths.  So the total number of spins is given by $N_S=N_U+N_D+2$. The spins are numbered in clockwise direction, i.e., starting from the spin in contact with the left bath followed by the upper branch spins and then the lower branch spins implying that the spins numbered $`1'$ and $`N_U+2'$ interact with the left and right bath respectively. The state or configuration of our system is described as:
	\begin{equation}\label{eta define}
		\eta=\{\sigma_1,\sigma_2,\ldots ,\sigma_{N_u+2},\ldots\sigma_{N_S}\},
	\end{equation}
	where $\sigma_i$ denotes the state of $i^{th}$ spin and $\sigma_i \in\{-1 ,1\}$.  The energy corresponding to a particular configuration is decided by the Hamiltonian of the system. For the cases that we study in this paper,  only the nearest neighbor spin--spin interaction are considered. Thus the Hamiltonian is given by
	\begin{align}\label{ising}
		H^{I}_S=-\sum_i J_i \sigma_i \sigma_{i+1},
	\end{align}
	where $H^{I}_S$ is the  spin--spin interaction term in the total Hamiltonian, $J_i$ is the interaction strength between the $i^{th}$ and $(i+1)^{th}$ spin and $\sigma_{N_S+1}=\sigma_{1}$. This means that the energy cost for flipping a spin $\sigma_i$ is,
	\begin{align}\label{energy_cost}
		\Delta E_{i}=2J_i\sigma_i \sigma_{i+1}+2J_{i-1}\sigma_i \sigma_{i-1},
	\end{align}
	subject to the periodic boundary conditions. Due to the interaction with the baths as well as the internal interaction between the spins, the configuration of the system may change resulting in the flow of heat currents in our system.
	
	\subsection{Time evolution dynamics }\label{subsec::time evolution,Classical Current Definitions}
	To carry our analysis forward, we first need to define the appropriate time evolution dynamics for our system. The master equation for our system is characterised by configuration pairs $(\eta,\eta^i)$ where $\eta^i$ is same as $\eta$ except for the flip of $i^{th}$ spin and is given as:
	\begin{align}\label{master_equation}
		\frac{d P(\eta)}{dt}=\sum_{i=1}^{Ns}r_{\eta^i\to\eta}P(\eta^i)-\sum_{i=1}^{Ns}r_{\eta\to\eta^i}P(\eta),
	\end{align}
	where $P(\eta)$ is the probability of finding the system in the configuration $\eta$ at time $t$ and $r_{\eta^i\to\eta}$ is the transition rate for flipping the spin $i$ and taking the system from configuration $\eta^i$ to $\eta$.   As already discussed above, for our model the nodal spins are in direct contact with the baths while the bulk spins are not. This suggests that to study our system we need hybrid dynamics with separate transition rates for nodal and bulk spins.
	
	Firstly,  because of their interaction with the baths, the dynamics of the nodal spins can be studied using the Metropolis algorithm, for which a spin flips according to the following transition rate \cite{group_circulation}
	\begin{align}\label{transition_rates}
		r_{\eta \to \eta^i}^{\alpha}= \text{min}(1, ~e ^{-{\beta_{\alpha} \Delta E_{i}}} ),
	\end{align}
	where $\beta_{\alpha}=1/{k_B T_{\alpha}}$, $\alpha\in\{L,R\}$, $i\in\{1,N_U+2\}$ and we work in units where $k_B=1$. Since the bulk spins are not in contact with any baths, a dynamics involving random numbers is not suitable for them and  a deterministic energy-conserving  dynamics is needed for understanding their time evolution. Such kind of dynamics generally fall under Cellular automaton \cite{wolfram, Q2R}.  We use two different types of cellular automaton for our problem, first the Q2R dynamics in the subsection \ref{subsec::Q2R dynamics with symmetric upper and lower branch interaction strength}, and then the CCA dynamics in \ref{subsec::CCA model}. We now discuss the current difinitions required for the classical analysis.
	
	\subsection{Heat Current Definitions}
	To define the heat current flowing out of a bath, we look at the nodal spin associated with it and identify all the configuration pairs $(\eta,\eta^i)$ which differ by its flip. The corresponding heat energy flow is then given as:
	\begin{align}\label{total_current_define}
		I_{\alpha}=\sum_{(\eta,\eta^i)} \Delta E_{i} ~( r_{\eta \to \eta^i}^{\alpha} P(\eta)-r_{\eta^i \to \eta}^{\alpha} P(\eta^i)),
	\end{align}
	where the sum runs over all the configuration pairs that differ by a flip of the $i^{th}$ spin with $\alpha\in\{L,R\}$ for $i\in\{1,N_U+2\}$. Since the bulk spins are not in contact with the baths, their flipping should cause no change in total energy of the system for any of the dynamics used. As a result, the Eq. \eqref{total_current_define} gives us zero when applied for flipping of bulk spins. However, the flip of bulk spins still leads to a transfer of energy from one spin to other over a bond \cite{cellular}. To better understand this, let us consider the interaction energy associated with a spin $i$,
	\begin{align}
		E_i= -J_i \sigma_i \sigma_{i+1}-J_{i-1} \sigma_i \sigma_{i-1}.
	\end{align} 
	The above expression contains the two bond energy terms associated with the spin $i$. Since this spin does not interact with the baths, it can flip only when the associated energy cost given in Eq. \eqref{energy_cost} is either zero or can be compensated by some other non-interacting internal energy of the system. In both these cases, the individual bond energies change, signifying a transfer of energy through the bonds and hence a heat current. Focusing on the $`i\text{--}i+1'$ bond, the energy transferred through it on the flipping of $i^{th} $ spin is:
	\begin{align}
		\Delta E_{i,i+1}=2 J_i \sigma_i \sigma_{i+1}.
	\end{align}
	To find the associated heat current, we identify all the configuration pair  transitions related by the flip of $i^{th}$ spin but focus only on the corresponding energy change on the {$`i\text{--}i+1'$} bond. This gives the following expression for  heat current passing through bulk spin  $i$
	\begin{align}\label{branch_current}
		I_{U(D)}=  \sum_{(\eta,\eta^i)} \Delta E_{i,i+1}(r_{\eta \to \eta^i} P(\eta)- r_{\eta^i \to \eta}P(\eta^i)),
	\end{align}
	where  $I_{U(D)}$ is the heat current in the upper (lower) branch depending on the position of the $i^{th}$ spin. Since, the flipping of the bulk spins is deterministic, it will either definitely happen or is completely forbidden and as a result $r_{\eta \to \eta^i} \in\{0,1\}$, unlike the stochastic flipping of the nodal spins. Finally, current magnification occurs when the heat current in one of the branches is larger than the current flowing between the system and the bath. According to the convention used in the current definitions above, we get current magnification if both the branch currents have the same sign. 

	{	\subsection{Information about Methodology }\label{Simulation abd Numerical}
		{ We now discuss the methodology \cite{methods} used to find results for the classical case.} All the  results shown in this manuscript are for the steady state.  The simulation for the classical case are performed by randomly selecting a spin and flipping it depending on a suitable dynamics. Here, we use a hybrid dynamics involving the metropolis algorithm for the nodal spins and the Q2R or CCA dynamics for the bulk spins. Whichever the case, only one spin is allowed to flip per time step. The system is allowed to relax to the steady state and the required quantities are then calculated. The corresponding numerical and analytical calculations are done by writing the Master equation and solving the transition matrix like the one given in Eq. \eqref{W_matrix}. Since, we are only interested in the steady state calculations, we require the eigenvector corresponding to the eigenvalue $`0'$ of the transition matrix. On applying the probability normalisation condition on this eigenvector, steady state probabilities can be obtained. These can then be used to calculate the currents analytically. 
		
		\par \textbf{Note on experimental parameter range}: In the units that we work, for both the classical and the quantum systems, the effect on system due to the interaction with the baths is quantified by an expression of the type $e^{J/T}$. This means that for any meaningful dynamics to take place, the  temperature of the baths should correspond to the thermal energy of the same order as $J$. The value of $J$ is typically decided by the experimental setup being studied. For example the experimental setup studied by Coldea et al. in \cite{experiment_ising}, the value of $J$ is in few mev which roughly corresponds to the thermal energy associated with temperatures of order $10$K.} 
	
	\begin{figure}[!b]
		\centering
		\includegraphics[width=0.4\textwidth,height=0.1\textheight]{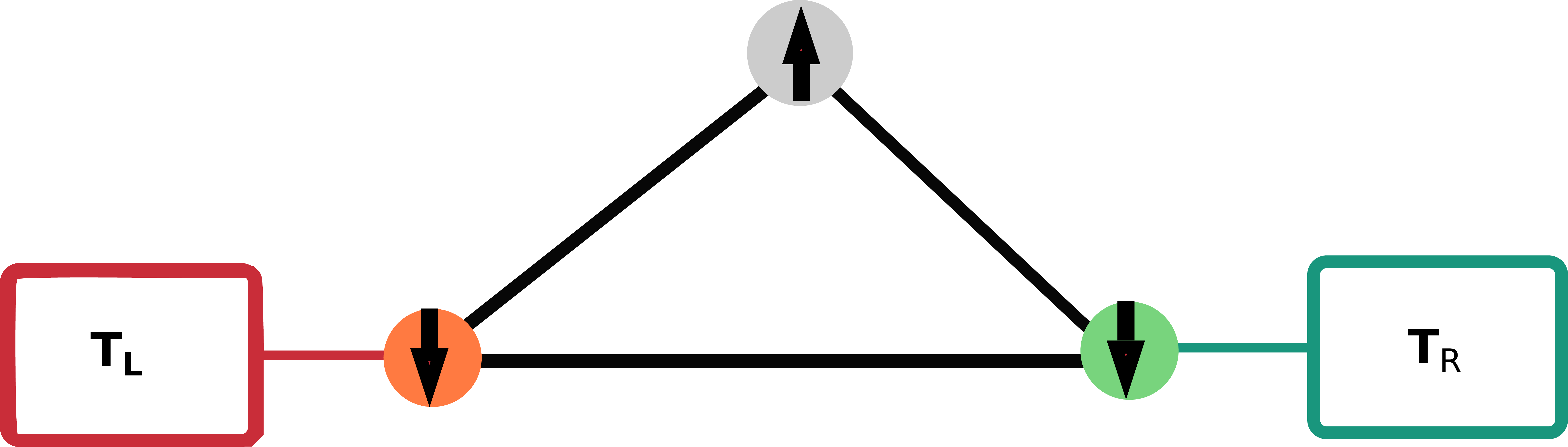}
		\caption{\label{model2} Illustration of the model system consisting of 3 spins with $N_U=1$ and $N_D=0$.}
	\end{figure}	
	\subsection{Q2R dynamics with symmetric upper and lower branch interaction
		strength}\label{subsec::Q2R dynamics with symmetric upper and lower branch interaction strength}
	We start our analysis by studying the Q2R dynamics \cite{cellular,CREUTZ198662} for the flipping of the bulk spins. This is one of the simplest model which allows us to have a deterministic mechanism for time evolution of the bulk spins while preserving the total energy of the system. The Hamiltonian for this model is the same as the typical Ising Hamiltonian and is obtained by substituting $J_i=J$ in equation \eqref{ising}. According to this dynamics, a bulk spin $i$ can flip only if its neighboring spins have opposite orientation i.e.,
	\begin{align}\label{q2r_dynamics}
		\sigma_{i-1}=-\sigma_{i+1}.
	\end{align}
	This ensures that the total energy  change of the system in Eq.   \eqref{energy_cost} is zero on flipping of the bulk spin.
	
	\begin{figure*}[t]
		\centering 
		\subfigure []
		{\includegraphics[width=0.4\linewidth,height=0.33\linewidth]{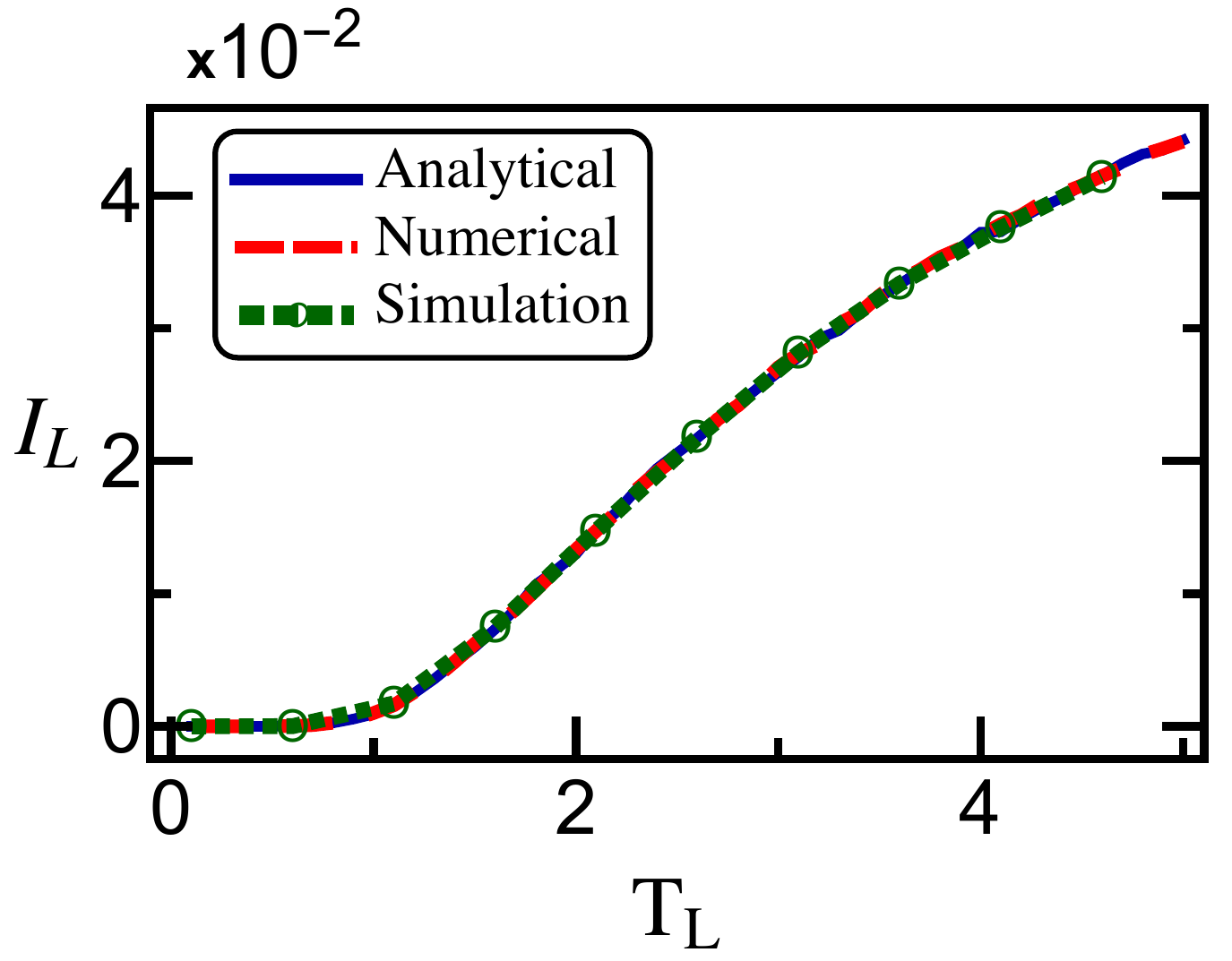}}
		\subfigure []
		{\includegraphics[width=0.4\linewidth,height=0.33\linewidth]{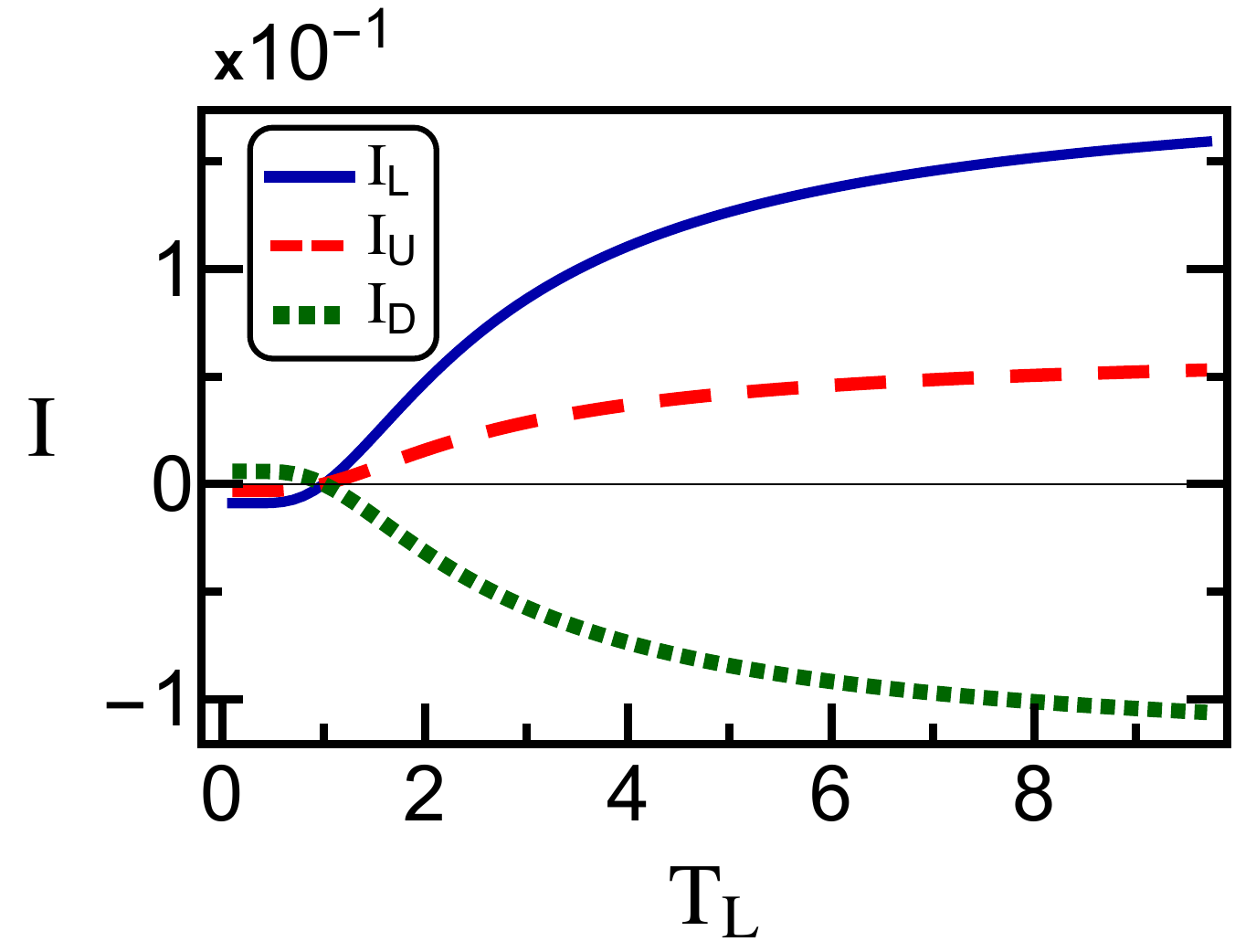}}
		\subfigure[]
		{\includegraphics[width=0.4\linewidth,height=0.33\linewidth]{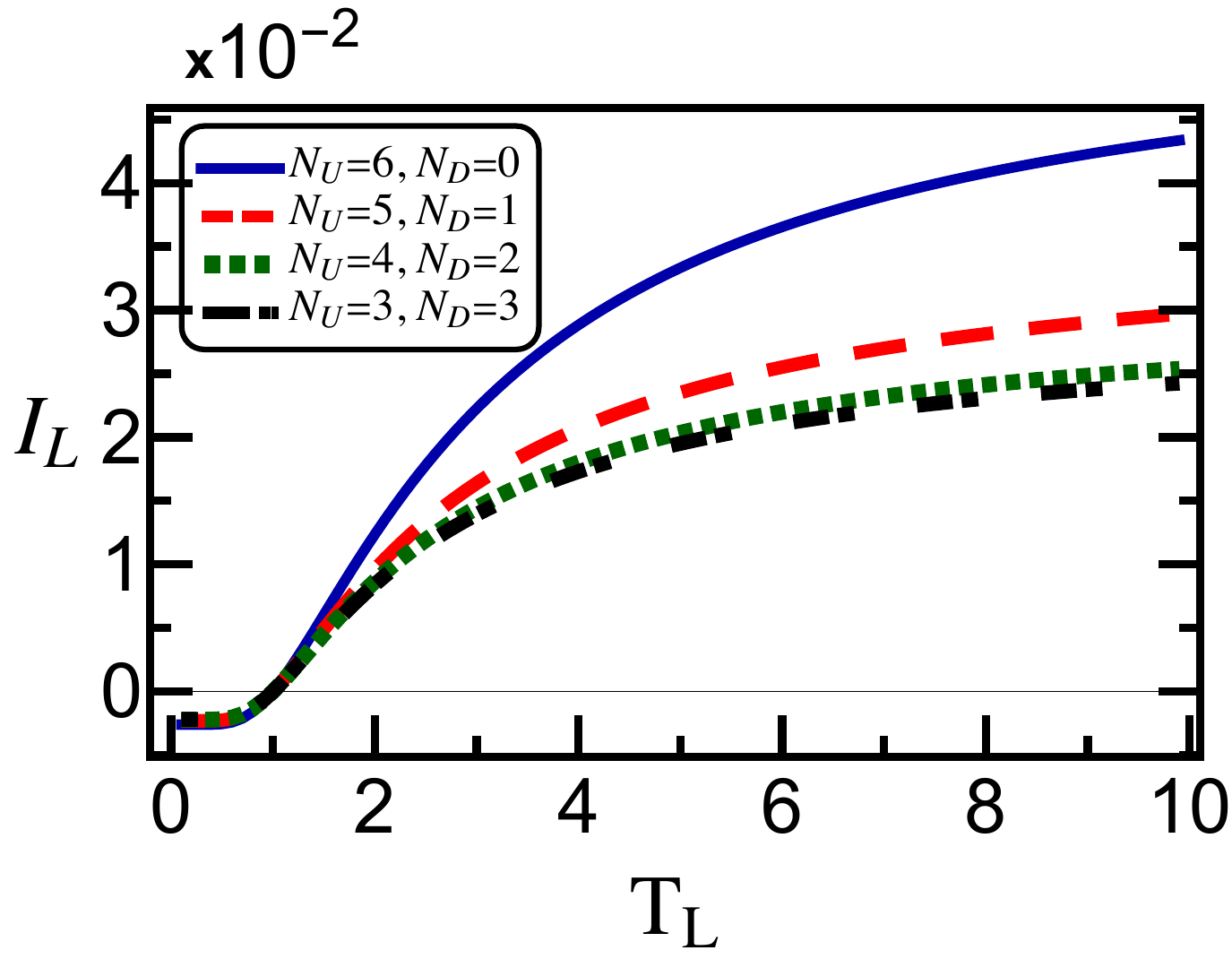}}
		\subfigure []
		{\includegraphics[width=0.4\linewidth,height=0.33\linewidth]{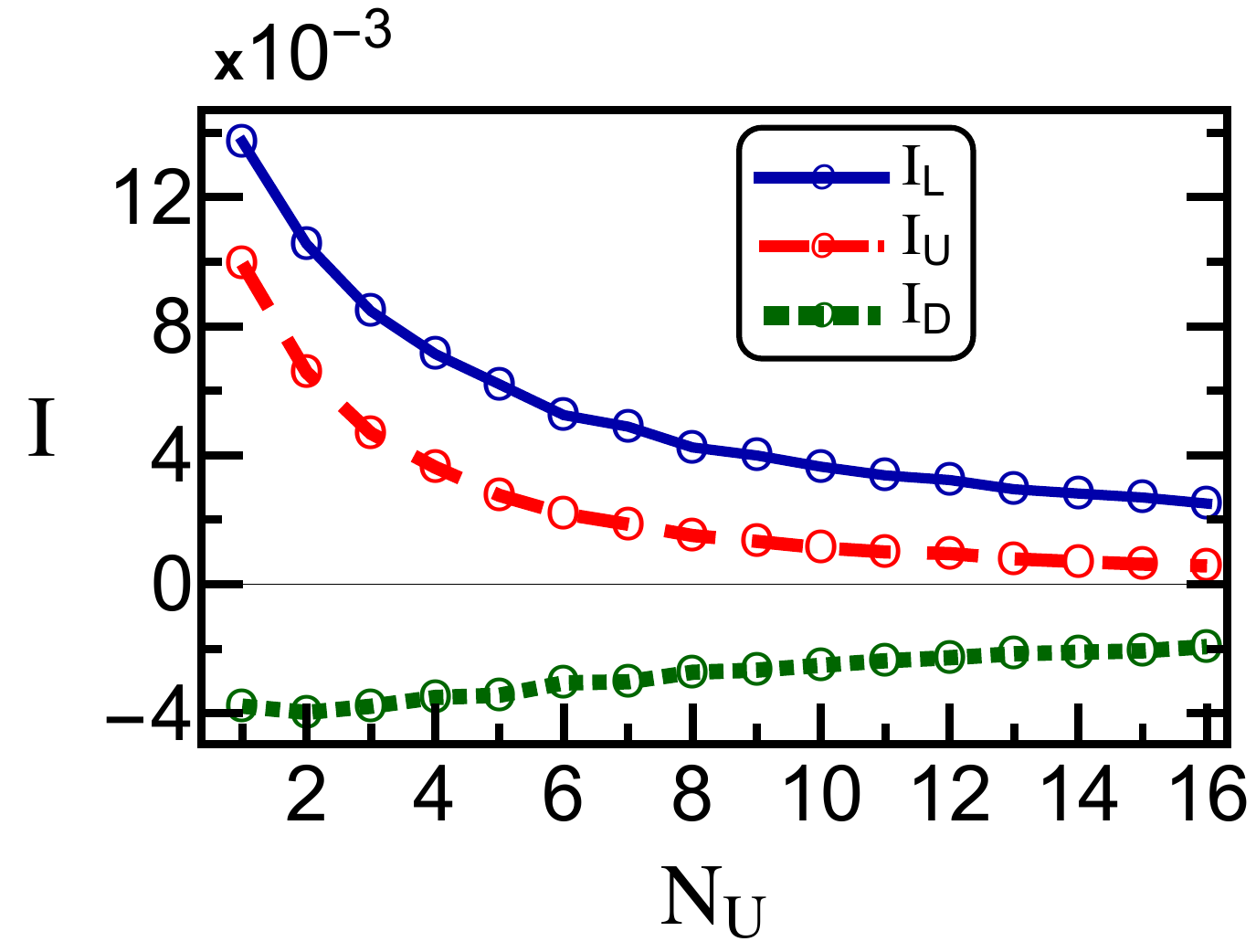}}
		
		\caption{(Color Online) \textbf{(a)} Comparison of total current ($I_L$) between  analytical, numerical  and simulation results  for the 3 spin system with $N_U=1$,  $N_D=0$, \textbf{(b)} Variation of Heat Currents ($I$) with $T_L$ for $N_U=1$, $N_D=0$,
			\textbf{(c)} Variation of $I_L$ with $T_L$ for a 8 spin system  but with different branch spin distributions, \textbf{(d)} Variation of Heat Currents ($I$) with $N_U$ for $N_D=4,T_L=2,T_R=0.1$. For all the cases unless otherwise specified,  we use $T_R=1$, $J=1$.}
		\label{symmetric_J_figures}
	\end{figure*} 
	To better understand the implications of the Q2R dynamics, we perform an analytical study for the minimalistic model with just three spins and $N_U=1$ and $N_D=0$ as shown in Fig. \ref{model2}. This being the simplest case, where the required asymmetric spin branching is possible, offers a good starting point. The master equation \eqref{master_equation} for this system written in the matrix form is given by,
      	\begin{widetext}
		\begin{align}\label{W_matrix}
			\frac{d}{dt} \begin{pmatrix}
				P(\downarrow,\downarrow,\downarrow)\\
				P(\uparrow,\downarrow,\downarrow)\\
				P(\downarrow,\uparrow,\downarrow)\\
				P(\uparrow,\uparrow,\downarrow)\\
				P(\downarrow,\downarrow,\uparrow)\\
				P(\uparrow,\downarrow,\uparrow)\\
				P(\downarrow,\uparrow,\uparrow)\\
				P(\uparrow,\uparrow,\uparrow)
			\end{pmatrix}=\begin{pmatrix}
				-e^{-4J\beta_L}-e^{-4J\beta_R}&1&0&0&1&0&0&0\\
				e^{-4J\beta_L}&-3&0&1&0&1&0&0\\
				0&0&-2&1&0&0&1&0\\
				0&1&1&-3&0&0&0&e^{-4J\beta_R}\\
				e^{-4J\beta_R}&0&0&0&-3&1&1&0\\
				0&1&0&0&1&-2&0&0\\
				0&0&1&0&1&0&-3&{e^{-4J\beta_L}}\\
				0&0&0&1&0&0&1& {-e^{-4J\beta_L}-e^{-4J\beta_R}}
			\end{pmatrix}\begin{pmatrix}
				P(\downarrow,\downarrow,\downarrow)\\
				P(\uparrow,\downarrow,\downarrow)\\
				P(\downarrow,\uparrow,\downarrow)\\
				P(\uparrow,\uparrow,\downarrow)\\
				P(\downarrow,\downarrow,\uparrow)\\
				P(\uparrow,\downarrow,\uparrow)\\
				P(\downarrow,\uparrow,\uparrow)\\
				P(\uparrow,\uparrow,\uparrow)
			\end{pmatrix}.
		\end{align}
		Solving the master equation (as discussed in Sec. \ref{Simulation abd Numerical}), we arrive at the following steady state probabilities:
	\end{widetext}

        \begin{align}\label{steady_state_probability}
&P(\downarrow,\downarrow,\downarrow)=\frac{e^{4J \beta_L+4J \beta_R}}{\mathcal{D}_1}, &&P(\uparrow,\downarrow,\downarrow)=\frac{3e^{4J \beta_L}+5 e^{4J \beta_R}}{8\mathcal{D}_1},\nonumber\\
&P(\downarrow,\uparrow,\downarrow)=\frac{e^{4J \beta_L}+ e^{4J \beta_R}}{2\mathcal{D}_1}, &&P(\uparrow,\uparrow,\downarrow)=\frac{5 e^{4J \beta_L}+ 3e^{4J \beta_R}}{8\mathcal{D}_1},\nonumber\\
&P(\downarrow,\downarrow,\uparrow)= \frac{5e^{4J \beta_L}+3 e^{4J \beta_R}}{8\mathcal{D}_1},
&&P(\uparrow,\downarrow,\uparrow)=\frac{e^{4J \beta_L}+ e^{4J \beta_R}}{2\mathcal{D}_1},\nonumber\\
&P(\downarrow,\uparrow,\uparrow)=\frac{3e^{4J \beta_L}+5 e^{4J \beta_R}}{8\mathcal{D}_1},
          &&P(\uparrow,\uparrow,\uparrow)=\frac{e^{4J \beta_L+4J \beta_R}}{\mathcal{D}_1}.
        \end{align}
where,
\begin{align}\label{denominator}
\mathcal{D}_1&= 2 e^{2 J(\beta_R +\beta_L)}\mathcal{D},\nonumber \\ ~\text{with}~
\mathcal{D}&=\left(3 \cosh(2 J (\beta_R - \beta_L)) ~+~ e^{2 J (\beta_L + \beta_R)}\right),
\end{align}
and the notation $P(\uparrow,\downarrow,\downarrow)$ specifies the probability of getting the  left spin up, middle spin down, and right spin down respectively. Using the above probabilities and the rates defined in Eq. \eqref{transition_rates},  the steady state heat current flowing out of the left bath in Eq. \eqref{total_current_define} becomes,
	{\begin{align}\label{left_current}
			I_L&=\frac{8}{3} J(e^{-4J \beta_L}P(\downarrow,\downarrow,\downarrow)-P(\uparrow,\downarrow,\downarrow))\nonumber \\
			&= J \sinh(2 J (\beta_R - \beta_L))/\mathcal{D}.
	\end{align}}
	Similarly, the  branch heat currents are found by using the Eq. \eqref{branch_current} for the bulk spins in upper and lower branch and have the following expressions,
	\begin{align}\label{upper_branch_current}
		I_U&= \frac{J}{3} \sinh(2 J (\beta_R - \beta_L))/\mathcal{D},\nonumber \\
		I_D&= -\frac{2J}{3} \sinh(2 J (\beta_R - \beta_L))/\mathcal{D}.
	\end{align}
	From the above expressions for total and branch currents, we see that  all the currents go to zero for equal temperatures of baths as expected. For different bath temperatures, the branch currents add up to give us the total current i.e. $I_L=I_U-I_D$. The minus sign comes because of the convention used for defining current direction in the system ($+$ve when current direction is left to right). Also, the magnitude of the total current is always greater than the magnitude of the branch currents for any values of system parameters, so heat current magnification is not possible for this system. { Interestingly, the heat current flowing through a branch is inversely proportional to the number of spin-spin bonds in that branch. This is similar to Ohm's law in an electric circuit with the spin-spin bond behaving analogous to a resistor. }
	\begin{figure*}[!htbp]
		\centering 
		\subfigure []
		{\includegraphics[width=0.4\linewidth,height=0.33\linewidth]{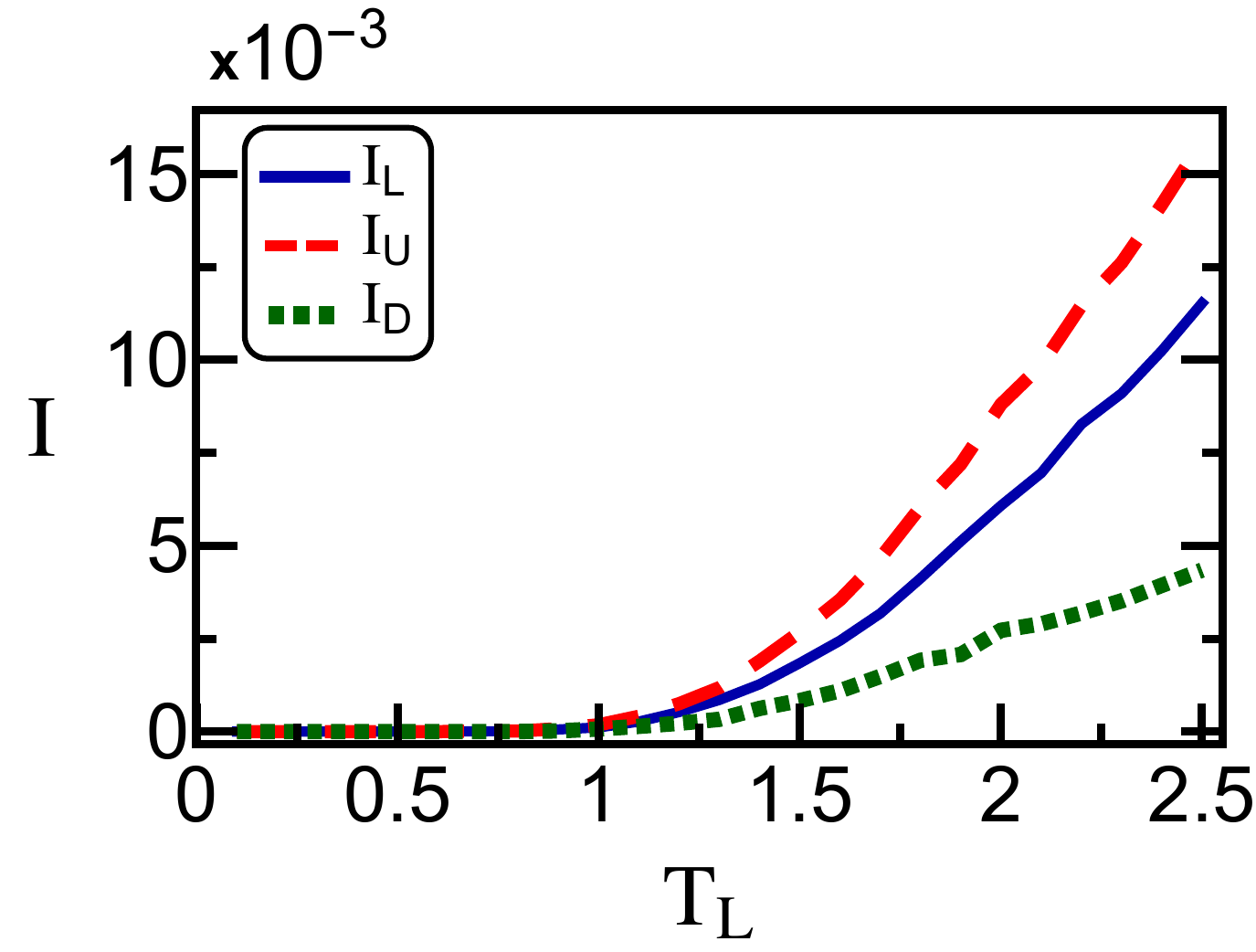}}
		\subfigure[]
		{\includegraphics[width=0.4\linewidth,height=0.33\linewidth]{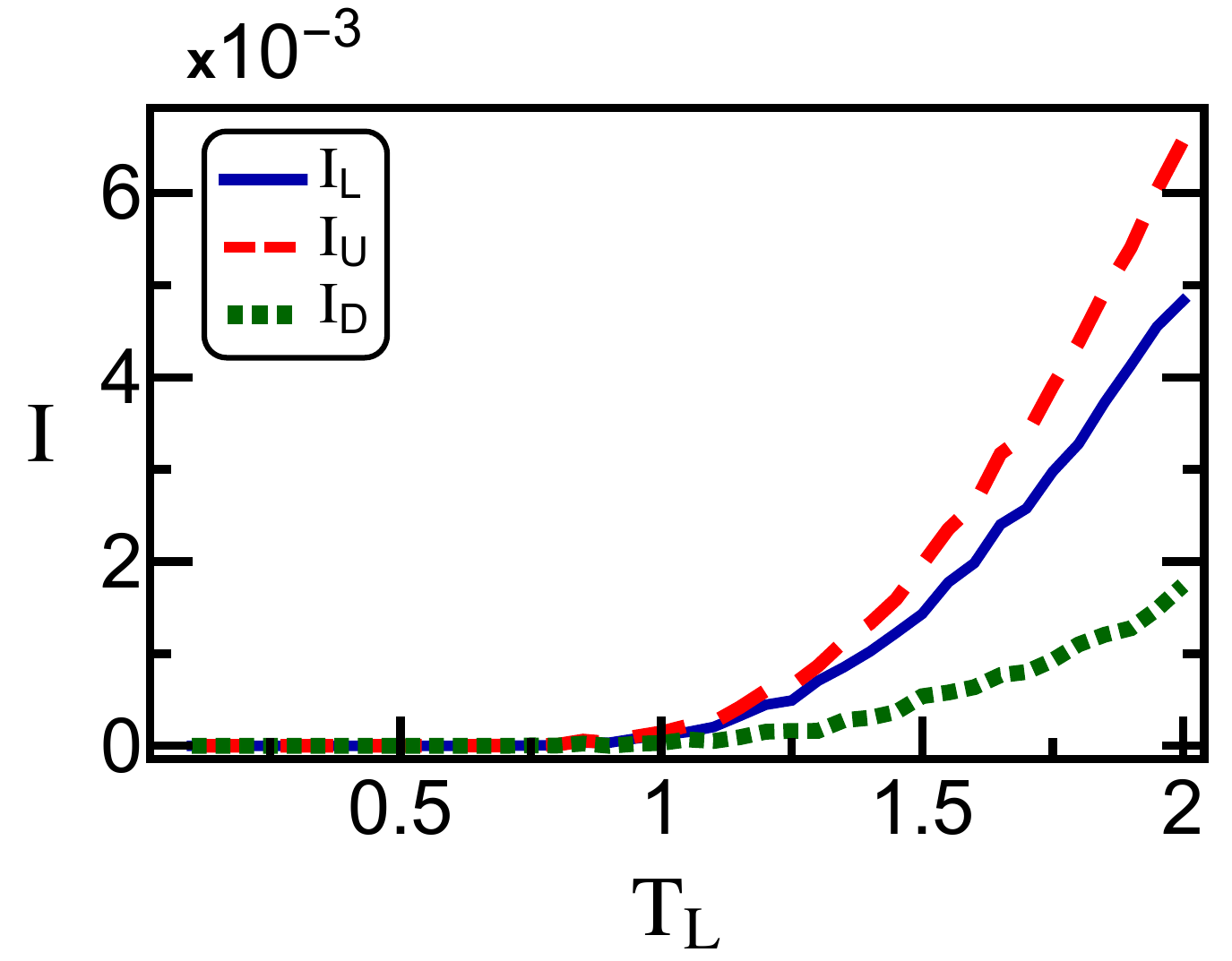}}
		\subfigure []
		{\includegraphics[width=0.4\linewidth,height=0.33\linewidth]{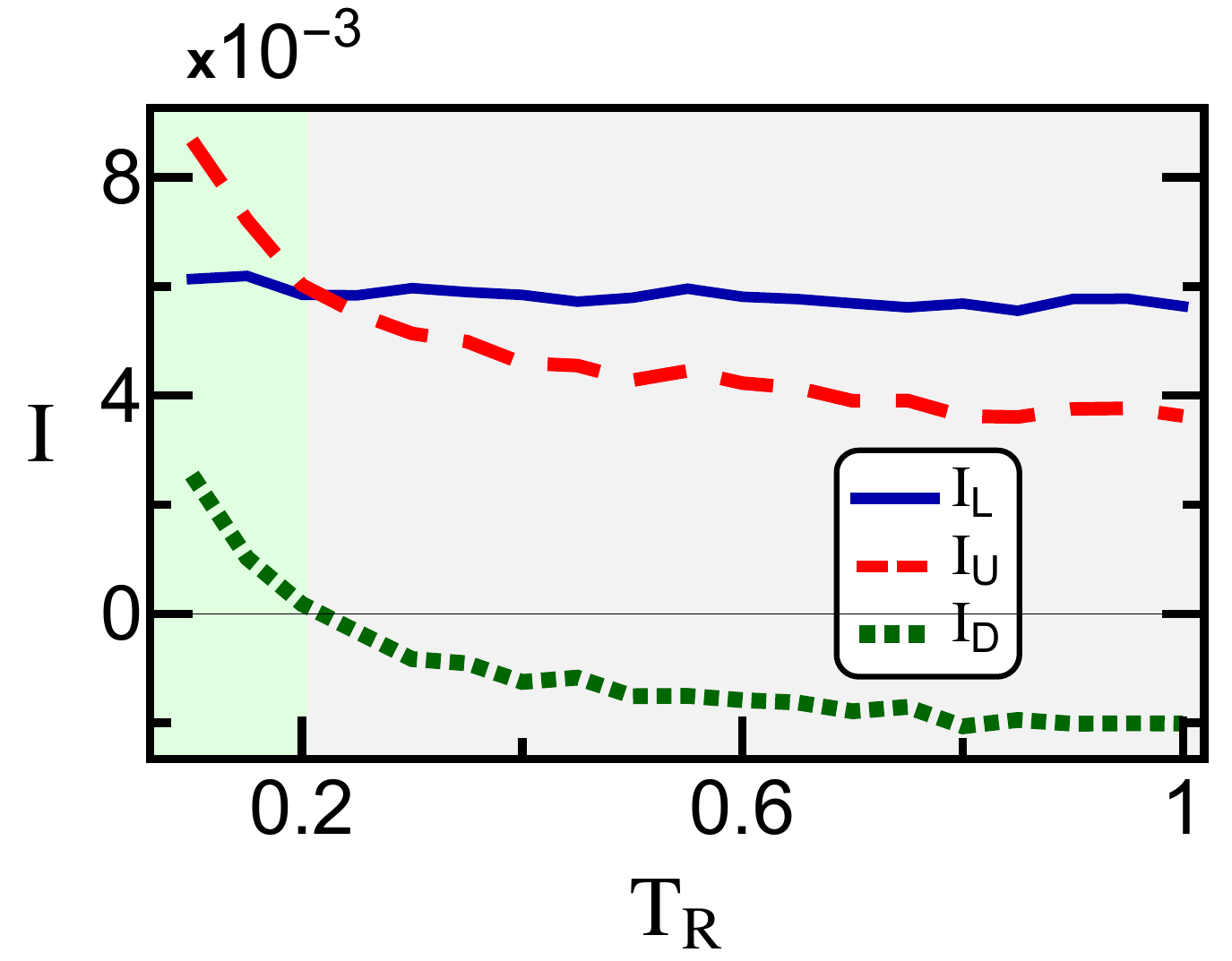}}
		\subfigure[]
		{\includegraphics[width=0.4\linewidth,height=0.33\linewidth]{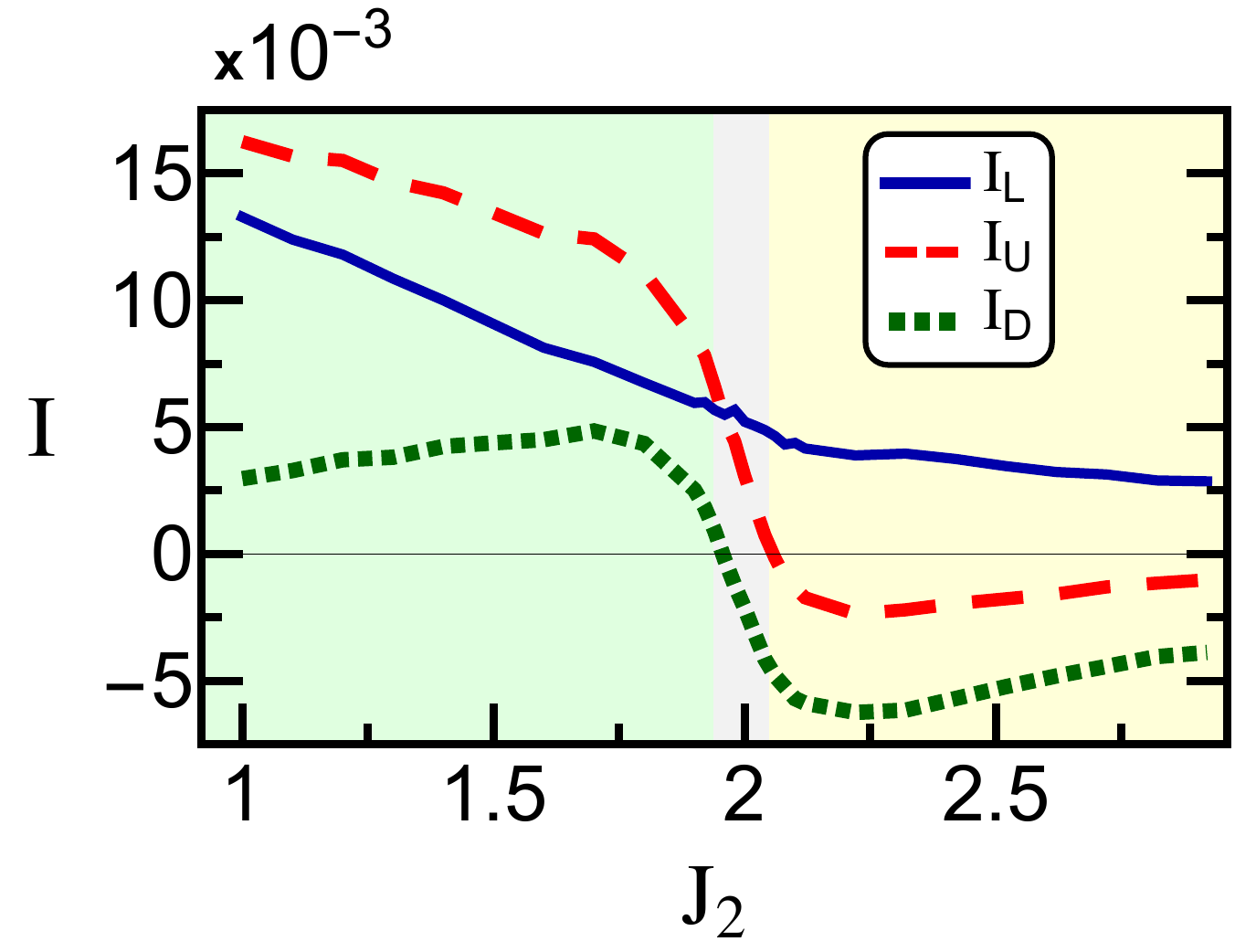}}
		\subfigure []
		{\includegraphics[width=0.4\linewidth,height=0.33\linewidth]{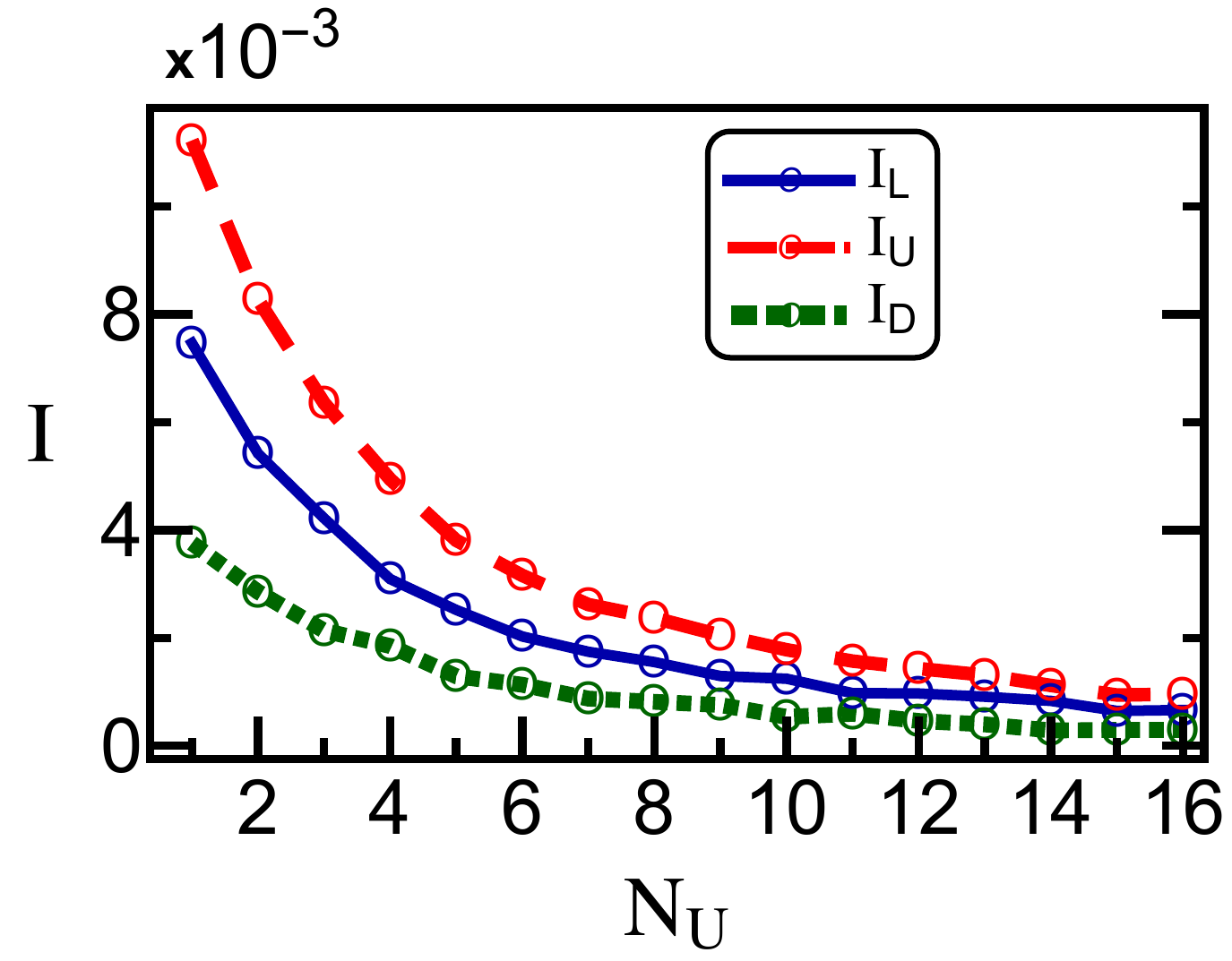}
		}
		\subfigure[]
		{\includegraphics[width=0.4\linewidth,height=0.33\linewidth]{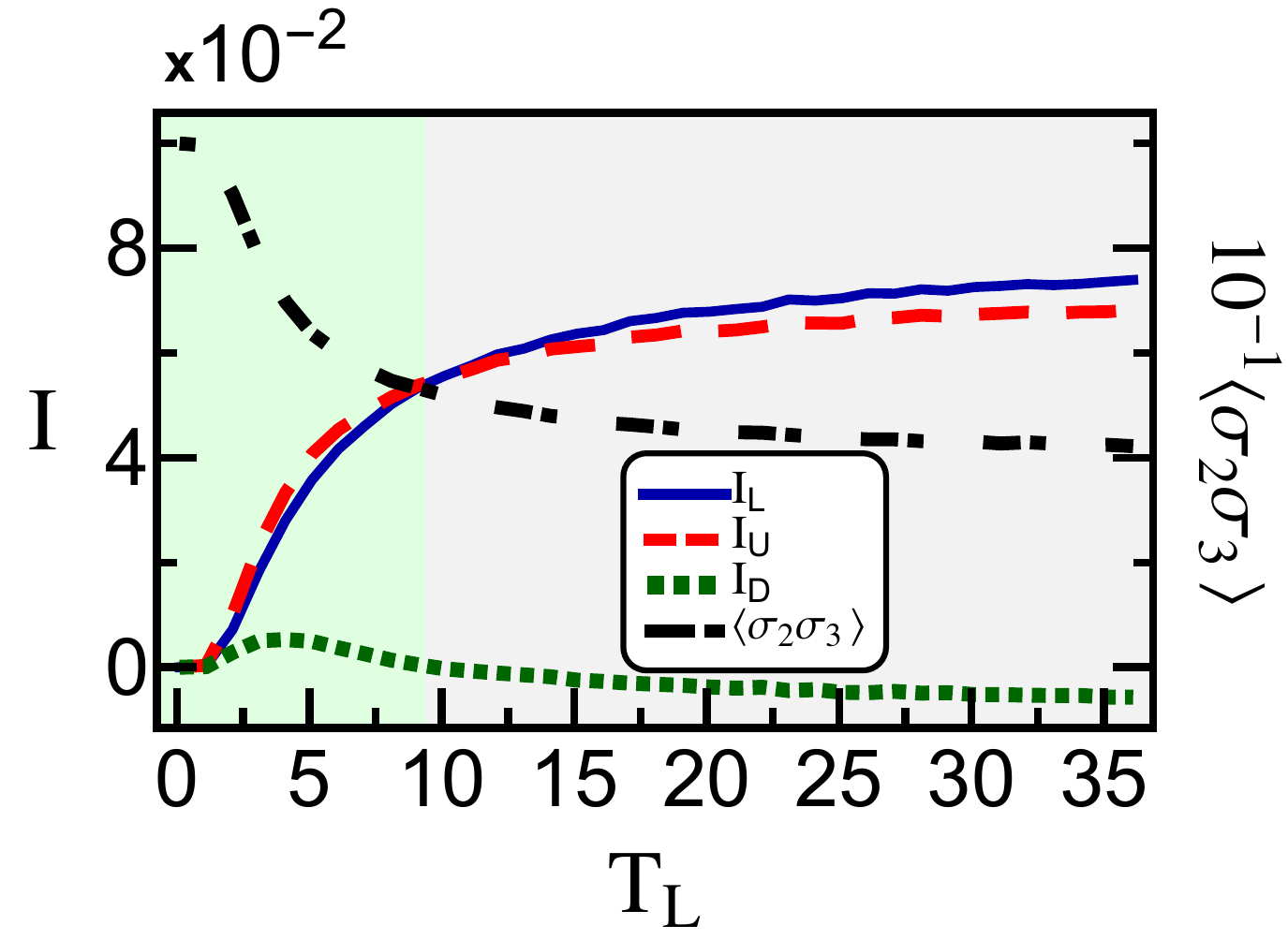}}
		\caption{ (Color Online) Variation of Heat Currents ($I$)  with \textbf{(a)} $T_L$,
			with \textbf{(b)} $T_L$ for   $N_D=2, N_U=3$, with \textbf{(c)}  $T_R$, with \textbf{(d)}  $J_2$, with \textbf{(e)}  $N_U$ for $N_D=4$ and \textbf{(f)} Variation of Heat Currents ($I$) and $\langle \sigma_2 \sigma_3 \rangle$ with $T_L$.
			For all the cases, unless otherwise specified we use $J_1=2, T_L=2, T_R=0.1, J_2=1.9, N_D=3, N_U=2$. 
			In all the figures where multiple regions of current circulation directions are present, we use green, gray and yellow background colors to indicate clockwise, parallel and anticlockwise circulating currents respectively. For all other cases, we use a white background.}
		\label{Asymmetric_Q2R_figures}
	\end{figure*}
	The corresponding simulation results for this model are  shown in Fig \ref{symmetric_J_figures}.  We see  in Fig. \ref{symmetric_J_figures}$(a)$ that for the $3$-spin system, our analytical and numerical results match perfectly with the simulation results. In Fig. \ref{symmetric_J_figures}$(b)$, we plot the current as a function of the temperature of the left bath $T_L$, keeping the temperature of right bath $`T_R'$ fixed . We find that the heat currents are non-monotonic functions of   temperature difference, initially increasing sharply and  then saturating for large temperature difference between the baths. { To understand why this is so, we put the limit of large  temperature gradient  $`T_L>>T_R'$ in  the equations \eqref{denominator}, \eqref{left_current} and \eqref{upper_branch_current}. Since we have fixed $T_R=1$, this means that $`T_L>>1'$ and $\beta_R \pm \beta_L \sim \beta_R$. Using this we get the following asymptomatic expressions for currents:
		\begin{align}\label{asymptotic limit}
			I_L&\sim \frac{J\sinh(2 J\beta_R)}{3 \cosh(2 J\beta_R  )+ e^{2 J\beta_R}},\nonumber \\
			I_U&\sim  \frac{\frac{J}{3}\sinh(2 J\beta_R)}{3 \cosh(2 J\beta_R  )+ e^{2 J\beta_R}},&&I_D\sim\frac{\frac{-2J}{3} \sinh(2 J\beta_R )}{3 \cosh(2 J\beta_R  )+ e^{2 J\beta_R}}.
		\end{align}
		We see that the above expressions are independent of $\beta_L (T_L)$, hence the current saturates for large values of $`T_L'$.} We also note from  Fig. \ref{symmetric_J_figures}$(c)$ that the total current flowing in the system does not only depend upon the total number of spins but also on how they are distributed in branches. In Fig. \ref{symmetric_J_figures}$(d)$ we see that the current decreases with an increase in the number of spins which is consistent with our earlier proposition of the spin-spin bonds behaving similarly to the resistors. From the above analysis, it is clear that we will not get current magnification just by branch spin number asymmetry for the Q2R dynamics and a dynamics with additional source of asymmetry is required. { We study one of such dynamics in the next section}



	\subsection{Q2R dynamics with asymmetric upper and lower branch interaction strength}\label{subsec::Q2R dynamics with asymmetric upper and lower branch interaction strength}
	Since, just the branch spin number asymmetry is not sufficient for generating current magnification. We now employ different spin-spin interaction strength in the upper and lower spin branch as a new source of asymmetry in our model. This effectively means that we have two different thermal wires for upper and lower branch. The Hamiltonian for this system is given as:
	\begin{align}
		H_S=-J_1 \sum_{i=1}^{Nu+1}\sigma_i \sigma_{i+1}- J_2\sum_{i=Nu+2}^{Ns}\sigma_i\sigma_{i+1}.
	\end{align}
	Similar to the previous case, it is possible to analytically solve this model for a $3$-spin system but the expressions are too complicated to be included here. However, we can still infer the general characteristic of this model by looking at the simulation results given in  Fig. \ref{Asymmetric_Q2R_figures}.
	We see in Fig. \ref{Asymmetric_Q2R_figures}$(a)$ that it is possible to get current magnification for asymmetric branch interaction strengths in the Q2R model if the system parameters are in optimal range. This deviates from our earlier observations and tells us that for different interaction strengths in upper and lower branch we can no longer assume the bond as being analogues to a resistor. In Fig. \ref{Asymmetric_Q2R_figures}$(b)$ we see that interchanging the number of spins in the branches still gives us {current circulation in the same direction but the magnitudes of currents change}. We also note (see Fig. \ref{Asymmetric_Q2R_figures}$(c)$) that we need the temperature of one of the heat baths to be an order below the energy scale of the bond interaction energy to achieve significant current magnification. Interestingly, as seen Fig. \ref{Asymmetric_Q2R_figures}$(d)$, increasing the difference between interaction strength of the branches does not necessarily increase the relative current magnification and a possibility for optimisation of current magnification exists. We see that the direction of current circulation changes depending on the relative magnitude of interaction strength in upper and lower branch. This may result in a scenario where no current flows through one spin branch and total current equals one of the branch currents. This happens for $J_2\sim 2$ in Fig. \ref{Asymmetric_Q2R_figures}$(d)$. We also see in  Fig. \ref{Asymmetric_Q2R_figures}$(e)$ that the magnitudes of the currents decrease with an increase in the number of spins and current magnification is possible even for same number of spins in the branches. To check how the current magnification depends on the spin-spin correlation defined as:
	\begin{align}
		\langle \sigma_i \sigma_j \rangle= \sum_{\{\sigma\}} P(\{\sigma\}) \sigma_i \sigma_j,
	\end{align}
	where the sum runs over all the system configurations $\{\sigma\}$ and $P(\sigma)$ is the probability of getting a particular configuration  $\sigma$. We plot the currents as well as the correlation between spin $2$ and $3$, namely $\langle \sigma_2 \sigma_3 \rangle$ in the Fig. \ref{Asymmetric_Q2R_figures}(f) and see that for the region where we get current magnification, the spin--spin correlation is high and when it decreases the magnification also decreases. We now discuss a possible physical mechanism behind this.
	
	\subsection{Possible Physical Mechanism}
	\begin{figure}[b] 
		\centering
		{\includegraphics[width=0.5\textwidth,height=0.25\textheight]{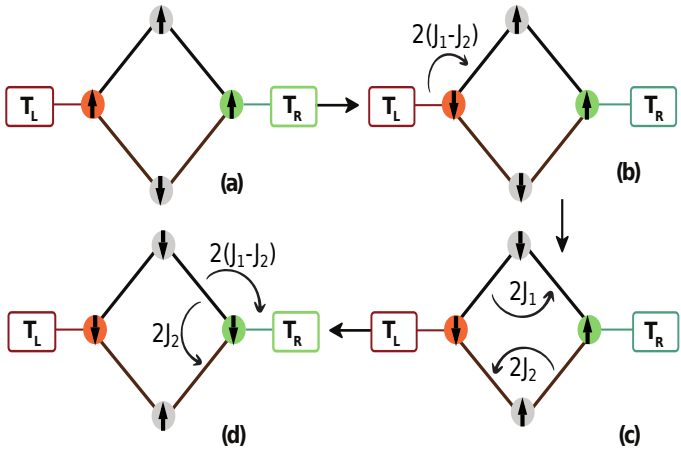}}
		\caption{Example of a process  which  result in  the transfer of energy $2(J_1-J_2)$ from left bath to the right bath.}
		\label{Explaination}
	\end{figure}
	To make sense of the results observed above, we study the example of an energy transfer process in a 4-spin system with $`N_U=1, N_D=1'$ as shown in Fig. \ref{Explaination}. The configuration change steps corresponding to this process are,
	\begin{align}
		\{\uparrow\uparrow\uparrow\downarrow\}\rightarrow\{\downarrow\uparrow\uparrow\downarrow\}\rightarrow\{\downarrow\downarrow\uparrow\downarrow\}\rightarrow\{\downarrow\downarrow\uparrow\uparrow\}\rightarrow\{\downarrow\downarrow\downarrow\uparrow\}
	\end{align}
	where as before the arrows indicate the states of spins numbered from left to right. The spins numbered $1$ and $3$ interact with the left and right baths respectively and follow Metropolis dynamics while spins $2$ and $4$ follow Q2R dynamics. The bond interaction strength is $J_1$ and $J_2$ for upper and lower spin branch respectively. The first step in the process involves the flipping of first spin via the Metropolis algorithm and signifies a transfer of energy $2(J_{1}-J_{2})$ from the left bath to the system (see Fig. \ref{Explaination}(a)). The next two steps in the process are shown in Fig. \ref{Explaination}(b) and involve the deterministic flipping of the bulk spins via the Q2R dynamics that result in energy transfer of magnitude $2J_1$ in the upper branch and $-2J_2$ in the lower branch successively. The final step involves the flipping of right nodal spin which indicates an energy transfer of magnitude $2(J_1-J_2)$ from the upper system branch to the right bath and simultaneously a energy transfer of $2J_2$ from the upper branch to the lower branch (see Fig. \ref{Explaination}(d)). Looking at the cumulative effect of all these steps, we get a process resulting in transfer of energy $2(J_1-J_2)$ from the left bath to the right bath accompanied by the flow of $2J_1$ energy in the upper branch and $2J_2$ energy in the lower branch, hence resulting in current magnification and clockwise current circulation inside the system. For the circulation of same energy in the anti-clockwise direction, we will have to transfer the energy $2(J_{1}-J_{2})$ from the right bath to the left bath which has less probability because of the lower temperature of the right bath, hence this process gives us a net current circulation. 
	
	\begin{figure}[t]
		\centering 
		\subfigure []
		{\includegraphics[width=0.49\linewidth,height=0.4\linewidth]{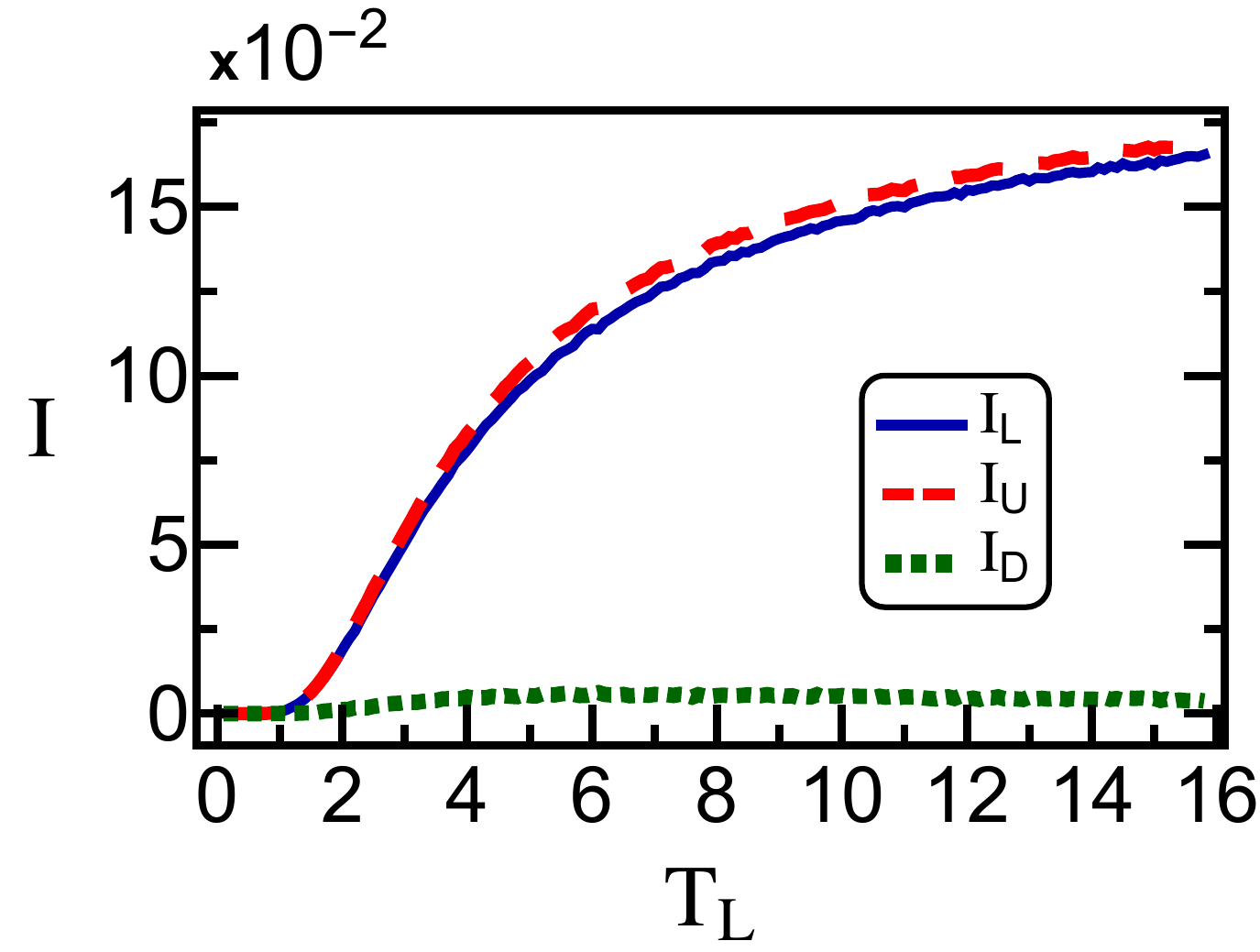}}
		\subfigure[]
		{\includegraphics[width=0.49\linewidth,height=0.4\linewidth]{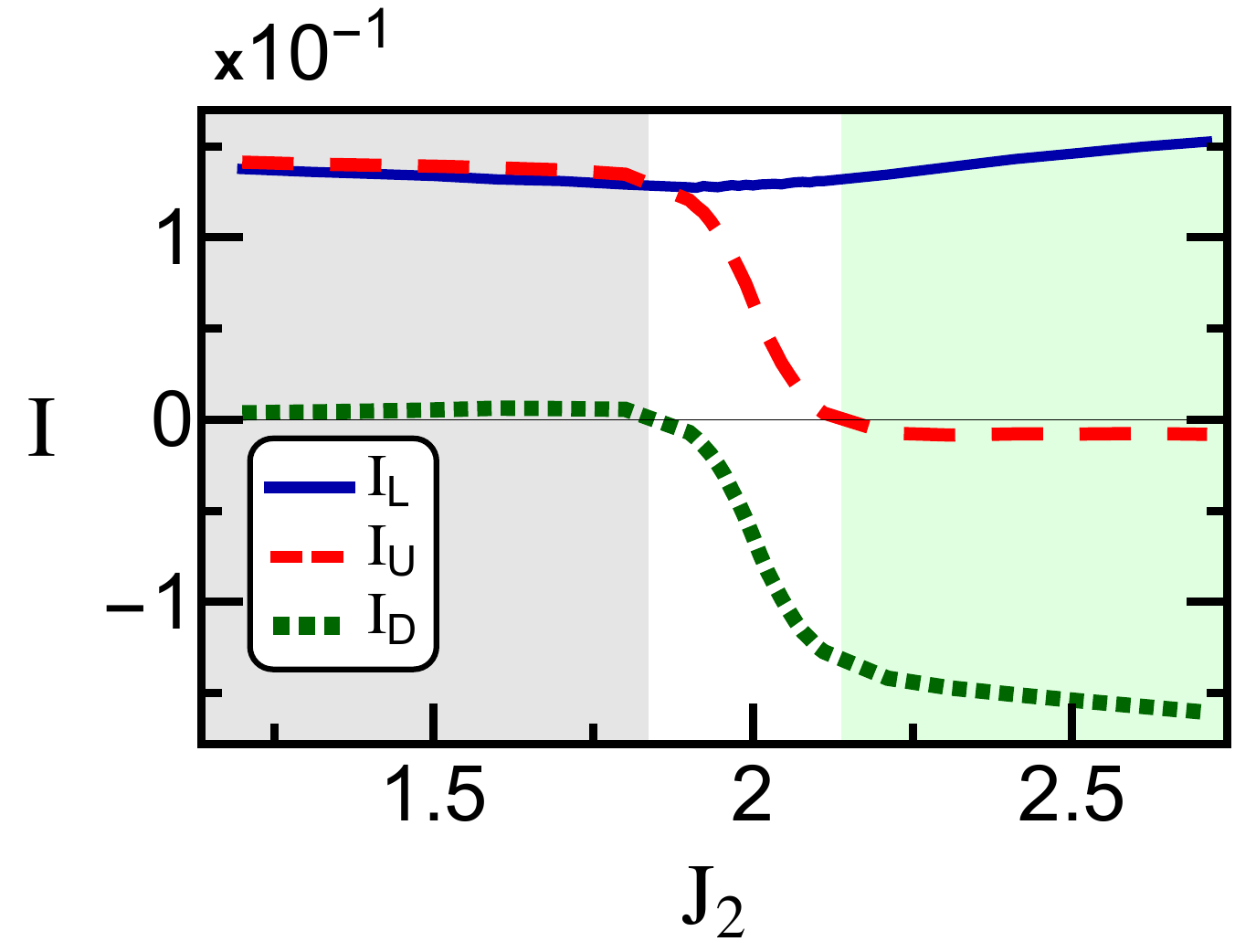}}
		\caption{(Color Online) Variation of Heat Currents ($I$)  with \textbf{(a)} $T_L$,
			with \textbf{(b)} $J_2$ for  the system shown in Fig. \ref{Explaination} with $N_D=1, N_U=1$,
			For all the cases, unless otherwise specified we use $J_1=2, T_L=2, T_R=0.1, J_2=1.9.$}
		\label{Classical_explanation_figures1}
	\end{figure}
	
	\begin{figure*}[t] 
		\centering 
		\subfigure []
		{\includegraphics[width=0.4\linewidth,height=0.33\linewidth]{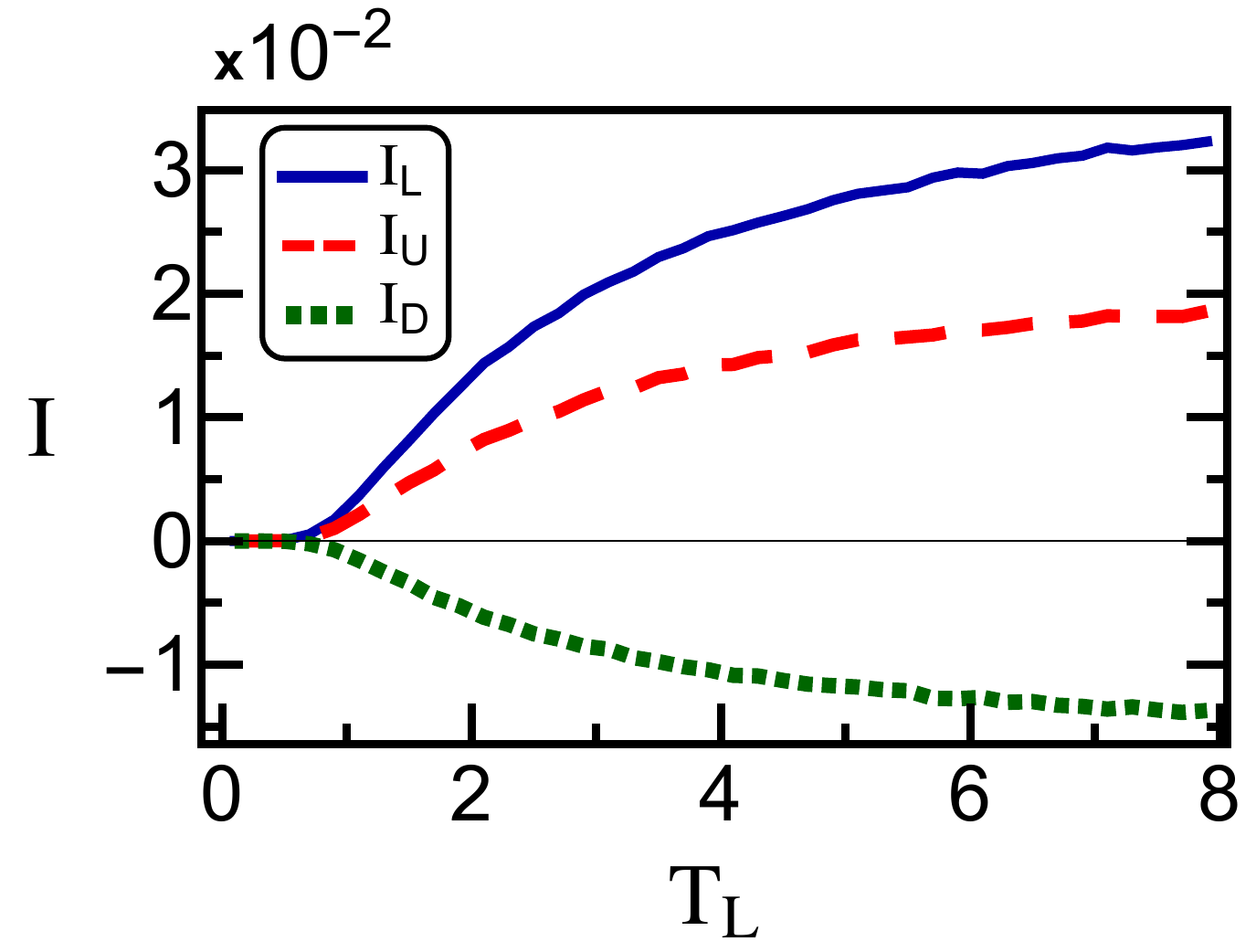}}
		\subfigure []
		{\includegraphics[width=0.4\linewidth,height=0.33\linewidth]{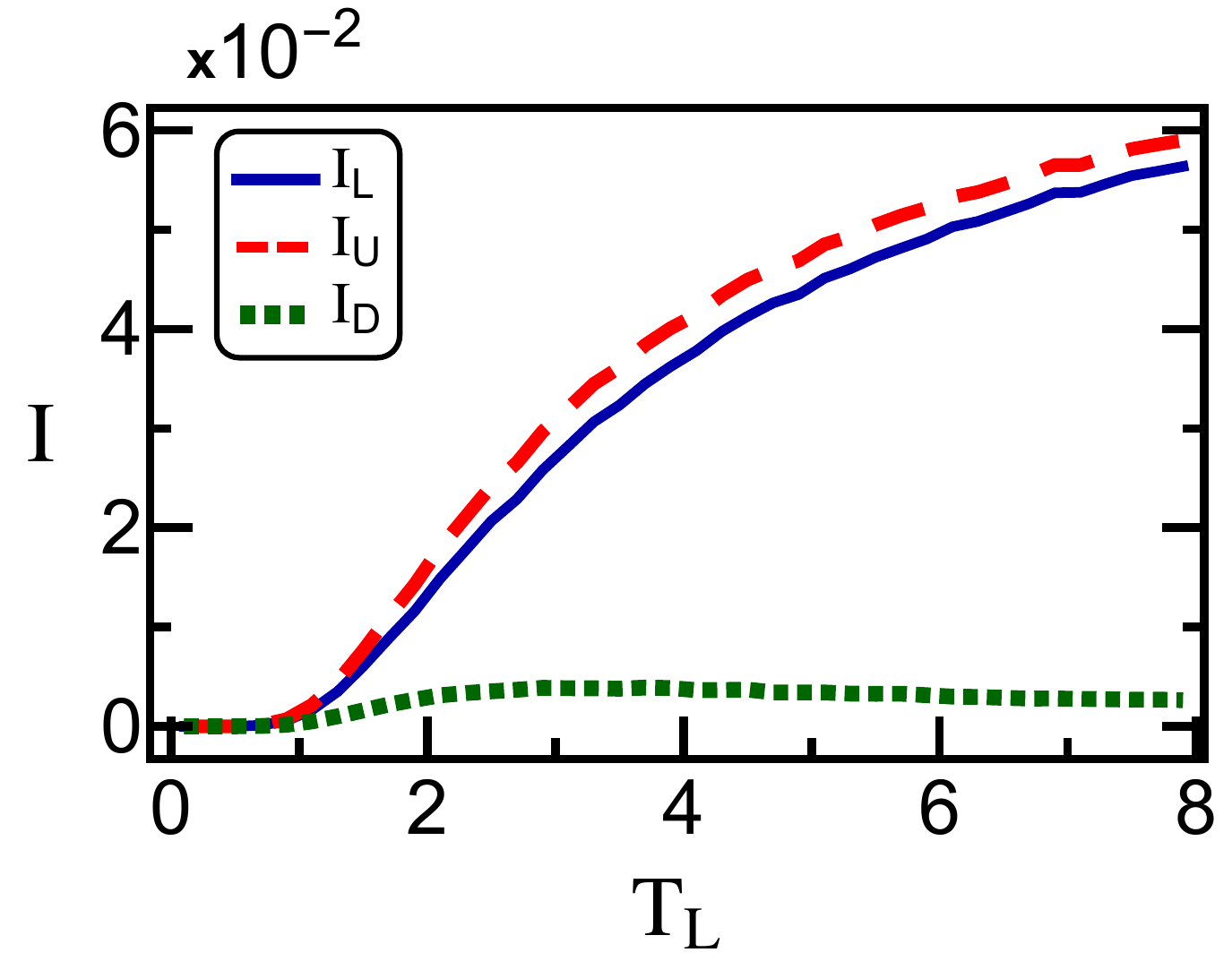}}
		\subfigure[]
		{\includegraphics[width=0.4\linewidth,height=0.33\linewidth]{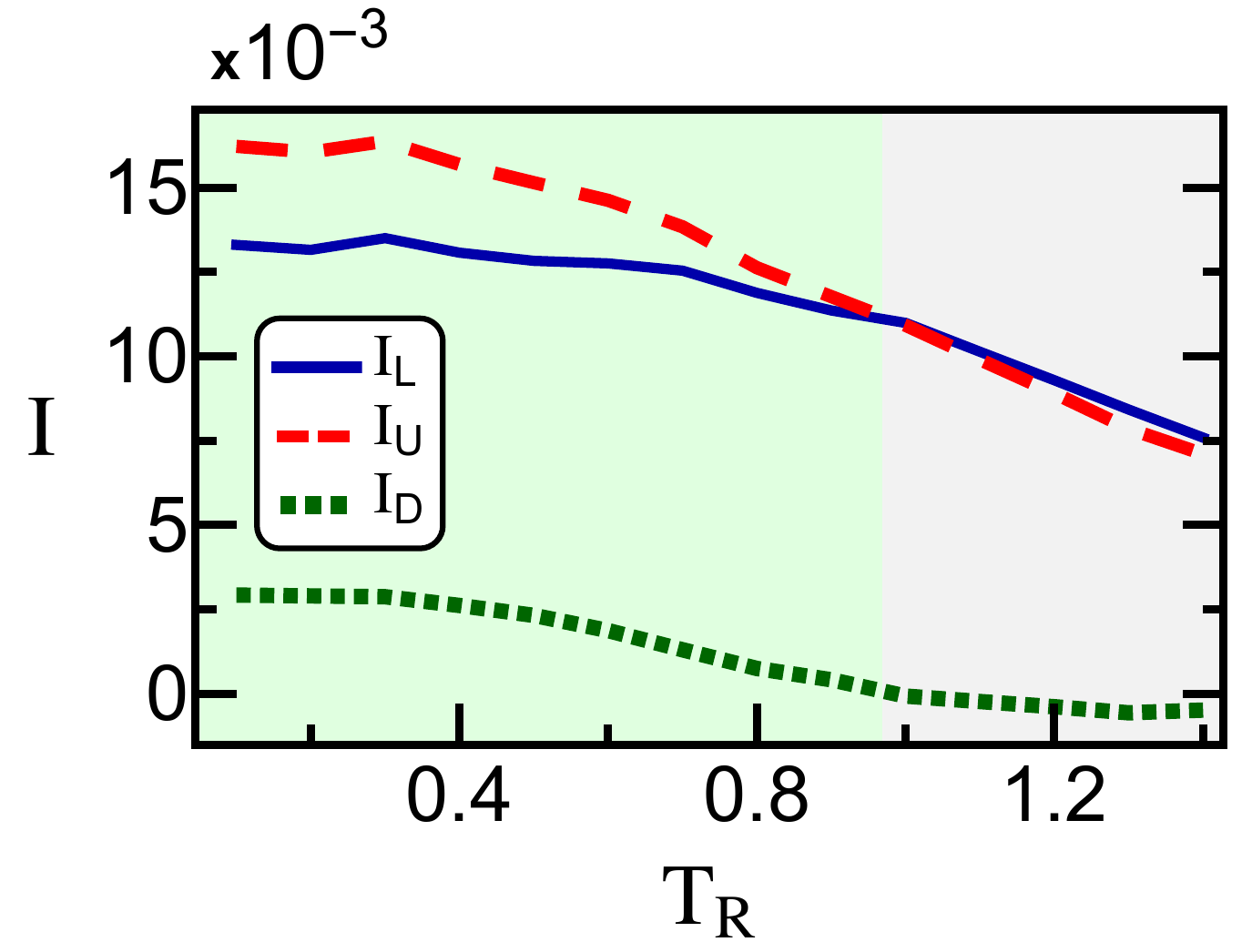}}
		\subfigure[]
		{\includegraphics[width=0.4\linewidth,height=0.33\linewidth]{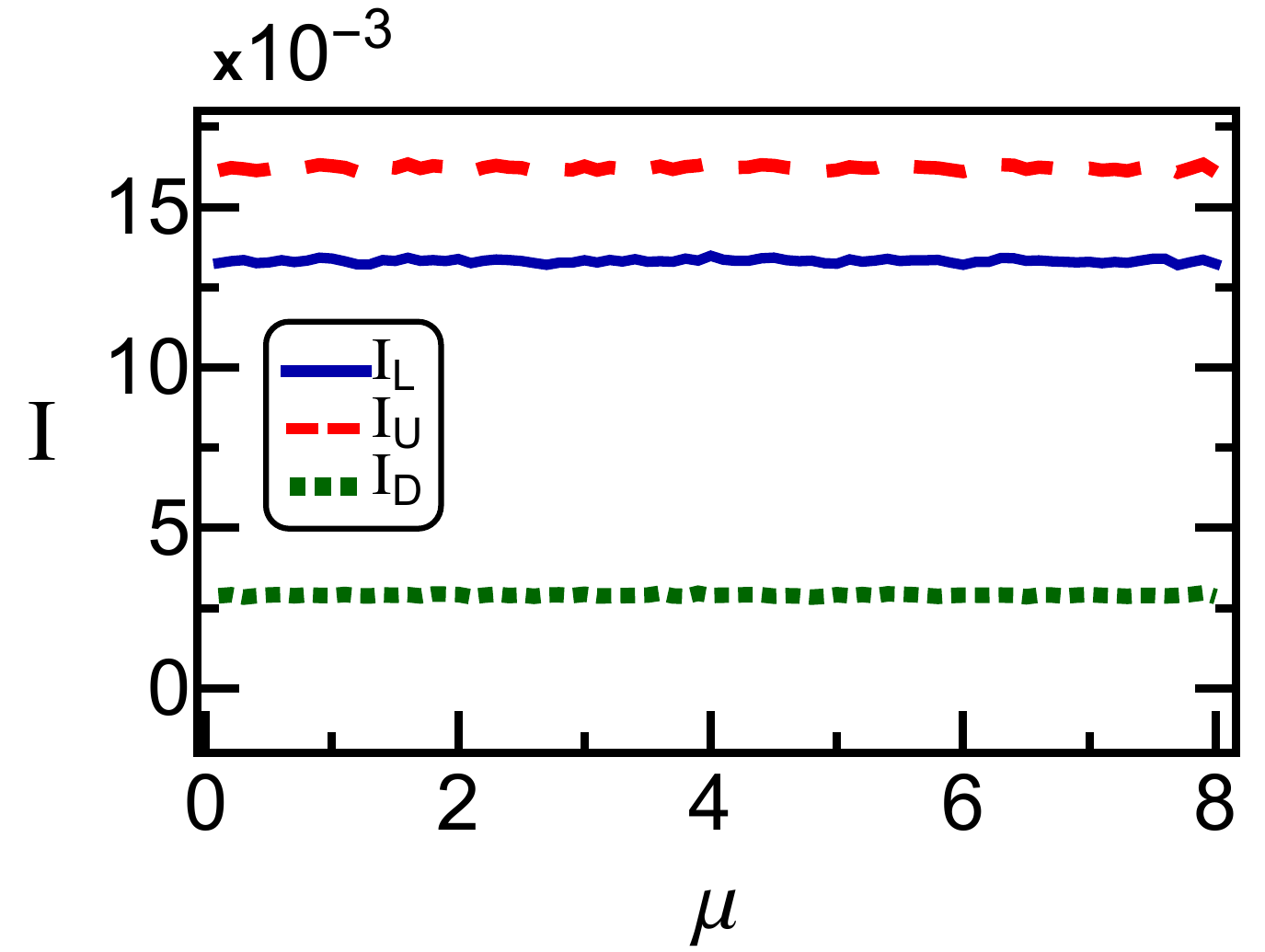}}
		\caption{(Color Online) Variation of Heat Currents ($I$) for the CCA model with \textbf{(a)} $T_L$ for  $J_1=J_2=1$, with \textbf{(b)} $T_L$, with \textbf{(c)}  $T_R$ and \textbf{(d)} with  $\mu$. 
			For all the cases unless otherwise specified $J_1=2, T_L=2, T_R=0.1, J_2=1, N_D=3, N_U=2$ and {$\mu=4$}.}
		\label{Momentum_term_figures}
	\end{figure*}
	The other way to transfer energy from left to the right bath includes a process where the energy transfer is of magnitude $2(J_1+J_2)$. This gives us no current circulation, but since the energy is of higher magnitude this process has less probability of occurrence. Thus overall the process with current circulation dominates the energy transfer and hence current magnification is happens on an average. But if the temperature of right bath is increased, the probability of the reverse process also increases and current circulation decreases as was seen in Fig. \ref{Asymmetric_Q2R_figures}$(b)$ . Increasing the gap between values of $J_1 $ and $J_2$ increases the energy transfer magnitude $2(J_{1}-J_{2})$ making the process less and less probabilistic, hence reducing relative current magnification. Therefore large current magnification is achieved if the interaction strengths are of similar order of magnitude (see Fig. \ref{Asymmetric_Q2R_figures}$(c)$ ). The above mechanism also explains why we will not get any heat current for symmetric interaction strength in both the branches as the net current due to this process becomes zero. No such heat transfer process is possible when we have the two nodal spins directly connected to each other, so there will be no current magnification in such a case as well. {The corresponding simulation results for the system shown in Fig. \ref{Explaination} are shown in Fig. \ref{Classical_explanation_figures1}. We see qualitatively similar results for this case as in Fig. \ref{Asymmetric_Q2R_figures}. Though, the value of the circulating current is small pertaining to the symmetry in the number of spin in upper and lower branch.} 
	
	\subsection{CCA Model} \label{subsec::CCA model}
	In the Q2R dynamics discussed above, the system Hamiltonian consisted of only the interaction energy terms and as such the heat current in the bulk could only be transferred through the passing of energy from one bond to the other. We now work with a model which allows additional ways for the system to possess energy. To achieve this we use the relatively complex CCA dynamics \cite{CREUTZ198662, cellular} for studying the time evolution of bulk spins. This dynamics allows for a change in interaction energy of the system by using an additional source of energy at each site, a local ``demon" \cite{CREUTZ198662}.  As stated earlier, for studying this dynamics we need to modify the total Hamiltonian and include an extra energy term to each spin site. This term corresponds to the presence of kinetic energy at each site and this results in the following total Hamiltonian:
	\begin{align}
		H_S=-J_1 \sum_{i=1}^{Nu+1}\sigma_i \sigma_{i+1}-J_2\sum_{i=Nu+2}^{Ns}\sigma_i\sigma_{i+1}+\mu\sum_{i=1}^{i=Ns} \tilde{\sigma}_i,\end{align}
	where $\mu$ is the scaling parameter for the kinetic energy reserve at each site and $\tilde{\sigma}_i$ represents the state of this new degree of freedom  with $\tilde{\sigma}_i \in \{0,1,2,3\}$. The presence of additional energy reserves at each site leads to the possibility of more bulk spins flips as some of the extra energy changes can be compensated. {Since the magnitude of energy supplied by the kinetic energy reserve is in multiples of $\mu$,  the bulk spin `$i$' can only flip under this dynamics if the  following condition is satisfied:
		\begin{align}\label{cca}
			\mu \tilde{\sigma}_i - \Delta E_i \in\{0, \mu, 2\mu, 3\mu\},
		\end{align}
		where,  `$\Delta E_i=2 J_i \sigma_i (\sigma_{i - 1} + \sigma_{i+1})$' and it can take the values `$\Delta E_U\in \{0,\pm 4 J_1\}$' or `$\Delta E_D \in \{0, \pm 4 J_2\}$' for upper and lower branch spins respectively. The above equation tells us that the bulk spin `$i$' can only flip if its kinetic energy reserve can compensate for the corresponding energy cost. So, the dynamics is still deterministic and total energy is conserved for the flip of a bulk spin.
A consequence of the above condition is that for CCA to allow additional flips than Q2R, the values `$\mu$' and `$J_1, J_2$' must be related. To illustrate this we give a few particular examples. If we choose, `$J_1=2, J_2=1$', and `$\mu=4$' then CCA dynamics can be applied for both the branches. Since, for $\mu=4$, $\mu\tilde{\sigma}\in\{0, 4, 8, 12\}$, $|\Delta E_{U}|\in\{0, 8\}$ and $|\Delta E_{D}|\in\{0, 4\}$. However, for some other value say $\mu=2$, we have $\mu\tilde{\sigma}\in\{0, 2, 4, 6\}$, $|\Delta E_{U}|\in\{0, 8\}$ and $|\Delta E_{D}|\in\{0, 4\}$. Hence, for the upper branch we essentially have the Q2R dynamics because the condition in Eq. \eqref{cca} is only satisfied for `$\Delta E_U=0$', but for the lower branch CCA dynamics still applies. We can see that irrespective of the particular values of $J_1$, $J_2$ and $\mu$, the above condition is always satisfied for `$0$' energy change and the Q2R dynamics always remains embedded in the CCA dynamics. In particular the Q2R dynamics studied in the earlier sections is a special case of the CCA dynamics for $\mu=0$.}  Since the  interaction energy of the spin bonds still remains the same, flipping of a spin signifies the same transfer of energy as before but that energy could come either from the neighboring bond or the kinetic energy reserve. The expressions for heat currents remain the same and are still given by equations \eqref{total_current_define} and \eqref{branch_current}. Simulating this model gives us the results shown in Fig. \ref{Momentum_term_figures}. \par In Fig. \ref{Momentum_term_figures}(a), we see that it is not possible to get current magnification even in the CCA model just by the branch spin number asymmetry and the total heat current distribution still has Ohm's law like characteristics for symmetric upper and lower branch interaction strength. Similar to the earlier cases, we see in Fig. \ref{Momentum_term_figures}(b) that we can get current magnification for this model if the spin--spin interaction strength differs in upper and lower branch. In Fig. \ref{Momentum_term_figures}(c), we see that the current magnification is possible only when the temperature of one of the baths is an order below the interaction energy of the bonds. {  Finally, on studying the variation of current with` $\mu$' in Fig. \ref{Momentum_term_figures}(d), we see that the steady state currents are independent of the values of $\mu$, even though CCA dynamics allows additional flips when $\mu \in \{2,4,8\}$. This means that we get similar steady state heat currents for both CCA and Q2R models for similar values of system parameters. The extra momentum term does not contribute critically to the steady state heat currents. This in agreement with what was observed by Saito et al. in \cite{cellular}. }
	\section{Quantum Analysis}\label{sec::Quantum}
	We now want to study the effects of branch spin number asymmetry on the heat current transport in the quantum domain. To do this, we perform a complementary study to the one performed by Xu et al. in \cite{main_reference}. It was shown by them that it is possible to get heat current magnification in a quantum system with $4$ spins interacting via modified Heisenberg exchange interaction if the onsite magnetic field is inhomogeneous and the total magnetisation of the system is conserved. However, it is difficult to realise this system experimentally as applying different magnetic fields at different sites for such small systems is in general a difficult task. So in place of the inhomogeneous magnetic field, we use unequal number of spins in branches similar to the classical models. This will also help us to distinguish the current magnification in quantum spin systems from the classical ones.
	\subsection{The model}
\begin{figure}[t]
		\centering
		\includegraphics[width=0.4\textwidth,height=0.15\textheight]{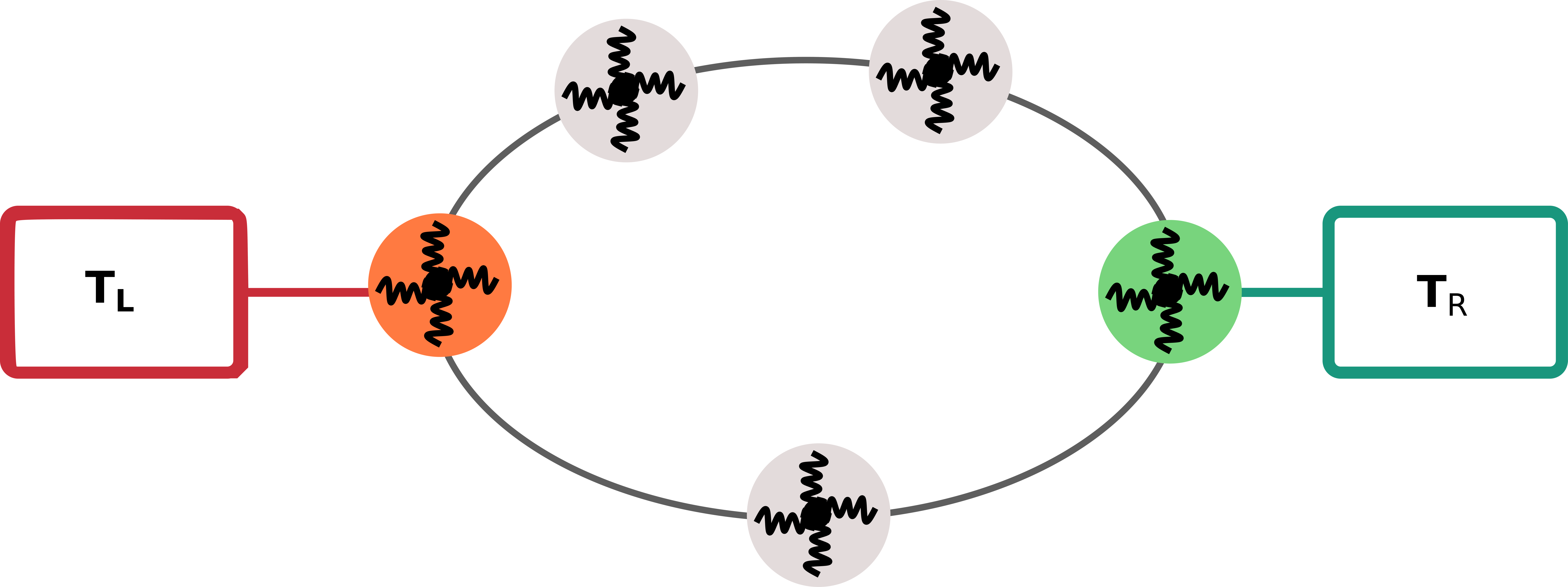}
		\caption{ Illustration of the Quantum Spin system consisting of 5 spins with $N_U=2$ and $N_D=1$.}
		\label{Quantum_Model}
	\end{figure}

	We study a five spin quantum system containing unequal branch spin numbers as shown in Fig \ref{Quantum_Model}. The spins are interacting with each other via a slightly modified Heisenberg XXZ interaction \cite{main_reference} and the system Hamiltonian is given as:
	\begin{align}\label{hamiltonian}
		\hat{H}_{S}&=J\sum_{i}(\hat{\sigma}^i_x\hat{\sigma}^{i+1}_x+\hat{\sigma}^i_y\hat{\sigma}^{i+1}_y)+\Delta \hat{\sigma}^i_{z'}\hat{\sigma}^{i+1}_{z'} ,
	\end{align}
	where, {$J$ is the interaction strength between the $x$ and $y$ components of neighboring spins,  $\Delta$ is called the asymmetry parameter of the Heisenberg XXZ interaction and signifies the interaction strength between $z$ components of the spins, $\hat{\sigma}_\alpha^i$ denotes the $\alpha\in\{x,y,z\}$ component of  the usual Pauli spin-$\frac{1}{2}$ matrix for the $i^{th}$ spin. The unit operators are denoted by $\hat{I}$ and}
	$\hat{\sigma}^i_{z'}=\frac{\hat{\sigma}^i_z+\hat{I}}{2}$. This slight modification to the usual XXZ interaction is done so that the transition to Fermionic system via the Jordan Wigner transformation \cite{coleman_2015} does not lead to a $\Delta$ dependent local chemical potential \cite{main_reference}. This system is interacting with two Bosonic baths with the following Hamiltonian,
	
	\begin{align}
		\hat{H}_{B_i}&=\sum_{n} \omega_n\hat{a}_n^{i\dagger}\hat{a}^i_n,
	\end{align} 
	where $\hat{a}_n^{i\dagger}(\hat{a}_n^{i})$ is the creation (destruction) operator of the $n^{th}$ mode of the `$i^{th}$'  bath. To achieve the spin branch  configuration given in Fig \ref{Quantum_Model}, the spins numbered `1' (`4') should interact with the left (right) baths respectively. { The system bath interaction is chosen to preserve initial magnetisation of the system and has the form: 
		\begin{align}
			\hat{H}_{SB_L}&= \hat{\sigma}^1_z\otimes \sqrt{\gamma} \sum_n c_n^L(\hat{a}^L_n+\hat{a}_n^{L\dagger}),\nonumber\\
			\hat{H}_{SB_R}&= \hat{\sigma}^4_z\otimes  \sqrt{\gamma} \sum_n c_n^R(\hat{a}^ R_n+\hat{a}_n^{R\dagger}),
	\end{align}}
	where, $\sqrt{\gamma}$ is the coupling strength between system and the baths and its value is always taken as positive.
	Though the operator $\hat{\sigma}_{z}$ can't change the magnetisation of a spin, it can still lead to a transfer of energy as the eigenstates of Hamiltonian and $\hat{\sigma}_{z}$ are not the same. 
	
	\subsection{Master Equation}
	To study the dynamics of the system, we again impose Markovian time evolution and hence the density matrix of our system satisfies a master equation of the type \cite{breuer2002},
	\begin{equation}\label{local}
		\frac{d}{dt}\hat{\rho}(t)=\mathcal{L}_{L}\hat{\rho}(t)+\mathcal{L}_{R}\hat{\rho}(t),
	\end{equation}
	where $`\hat{\rho}(t)'$ is the density matrix of the system of interest at time `t' and $`\mathcal{L}_{L(R)}'$ is the Liouvillian operator of the left (right) bath.  
	Specifically, for the analysis in this manuscript, the Redfield master equation \cite{Cattaneo_2019,breuer2002} is used.  This is because the more simpler Lindblad master equation doesn't give the correct coherences in the steady state and hence { we obtain  incorrect  branch currents in that case \cite{correction1,main_reference}}. The Redfield master equation is { used for the weak coupling regime between the system and the baths and is correct only upto order $\gamma$ \cite{Cattaneo_2019,breuer2002}. Its form for our model in the Schr$\ddot{o}$dinger picture and in the energy basis is given as  (working in the units where $\hbar=1$)},		
	\begin{align}\label{Redfield}
		\dot {\hat{\rho}}_{nm}=&-i\Delta_{nm} \hat{\rho}_{nm}+\gamma \sum_{i,j}\big{(}\hat{S}^{L}_{ni}\hat{S}^{L}_{jm}[\tilde{W}^{L}_{jm}+W^{L}_{ni}] \nonumber \\ 
		&-\delta_{mj}\sum_l \hat{S}^{L}_{nl}\hat{S}^{L}_{li}W^{L}_{li}(0)-\delta_{ni}\sum_l \hat{S}^{L}_{jl}\hat{S}^{L}_{lm}\tilde{W}^{L}_{jl}(0)\big{)}\rho_{ij}\nonumber\\&\textbf{+Right Bath Terms}
	\end{align}
	{The above equation can be divided into two parts, the first part not multiplied by $\gamma$ is just the typical time evolution of an isolated quantum system with the Schr$\ddot{o}$dinger equation and the part multiplied by $\gamma$ corresponds to the time evolution of the system  due to interaction with the baths.} Here $ \hat{\rho}_{nm}$ is the matrix element of the system density matrix and  $`\Delta_{nm}'$ is the energy difference between the $n^{th}$ and $m^{th}$ energy levels or $\Delta_{nm}=E_n-E_m$.
	$\hat{S}^{L}(\hat{S}^R)$ is the operator of the system that interacts with the left (right) bath and $\hat{S}^{L,R}=\hat{\sigma}^{1,4}_z$ for our case. Finally, the $W$ matrices depend on the system energy spectrum and the heat bath correlations. They are defined as:
	\begin{align}\label{w}
		W^{L}_{ij}=\int_{0}^{\infty}dt e^{-i\Delta_{ij}t} C^{L}(t),
	\end{align}
	\begin{align}
		\tilde{W}^{L}_{ij}=\int_{0}^{\infty}dt e^{-i\Delta_{ij}t} C^{L}(-t),
	\end{align}
	with $C^{L}(t)=Tr_B\text{ }(\hat{B}_L(t)\hat{B}_L \hat{\rho}_B(0))$, $\hat{B}_{L}=\sum_n c_n^L(\hat{a}_n+\hat{a}_n^\dagger)$ and $\hat{\rho}_B(0)$ is the equilibrium density matrix corresponding to the bath. {Ignoring the Lamb shift terms \cite{Cattaneo_2019,breuer2002}, $\tilde{W}^{L(R)}_{j,i}=W^{L(R)}_{i,j}$ and we can write}, 
	\begin{align}
		W^{L(R)}_{i,j}=
		\begin{cases}
			J(\Delta_{ij})N(\Delta_{ij},T_{L(R)}) & \Delta_{ij}>0, \\
			J(|\Delta_{ij}|)\big{(}1+N(|\Delta_{ij}|,T_{L(R)})\big{)}  & \Delta_{ij}<0,
		\end{cases}
	\end{align}
	where,  $N(\omega,T_{L(R)})$ is the Bose-Einstein distribution function,
	\begin{align}\label{eq:bose-einstein}
		N(\omega,T_{L(R)})=\frac{1}{e^{\omega/T_{L(R)}}-1}.
	\end{align}
	and we choose,
	\begin{equation}
		J(\omega) =\omega e^{-{\omega}/{\Omega_C}},
	\end{equation}
	as the spectral density of the baths. We fix $\Omega_C=10$ without loss of generality. { Since, the above equation is only correct upto the order $\gamma$, it means that $`\gamma'$ should be smaller than $`1'$ for any meaningful discussion, so we  take $\sqrt{\gamma}=0.1$ without loss of generality. The functional form of the bath spectral density and numerical values of $\Omega_C$ and $\gamma$ are similar to the ones taken in \cite{main_reference}.}
	\begin{figure*}[t]
		\subfigure []
		{\includegraphics[width=0.4\linewidth,height=0.33\linewidth]{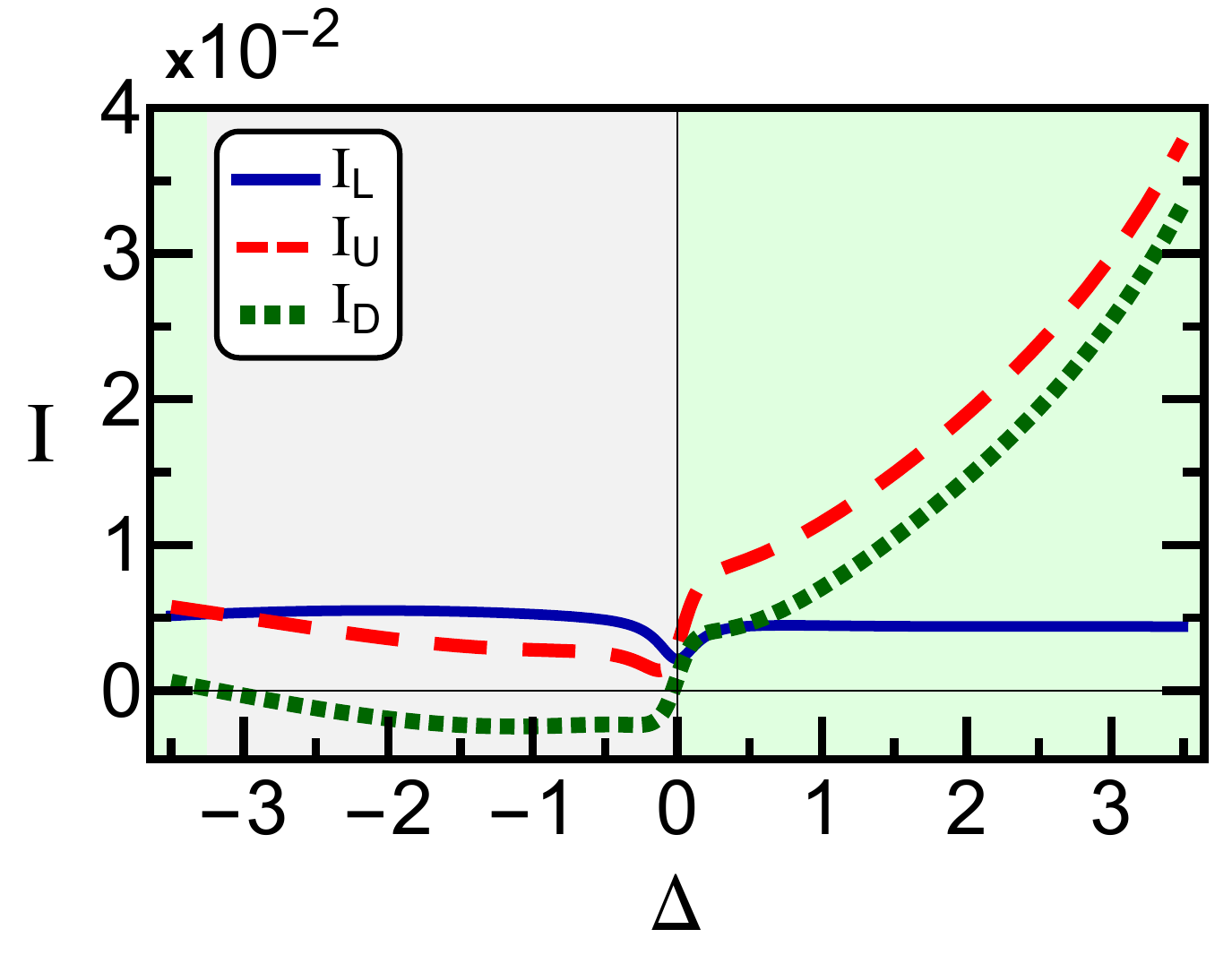}}
		\subfigure []
		{\includegraphics[width=0.4\linewidth,height=0.33\linewidth]{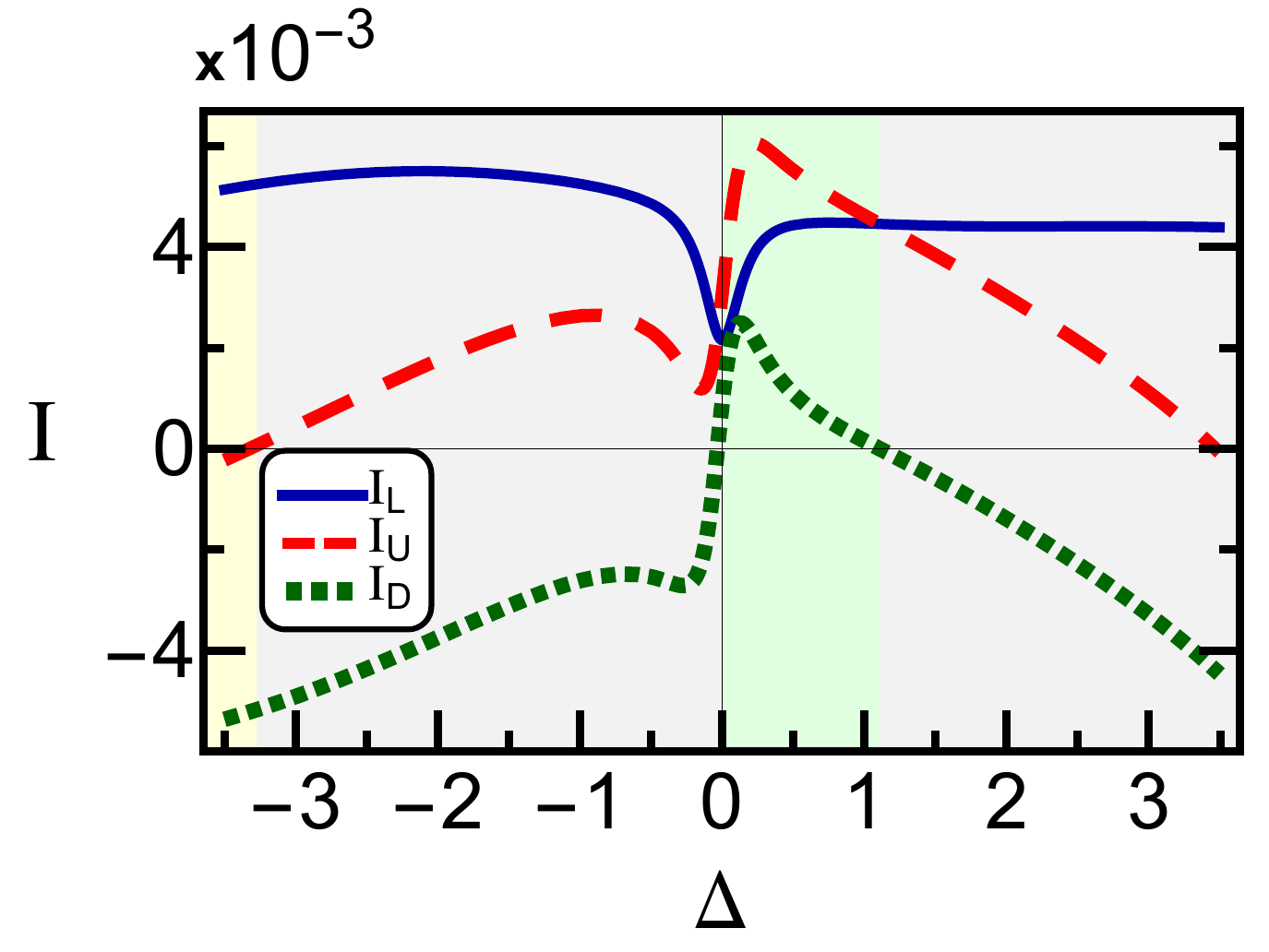}}
		\subfigure []
		{\includegraphics[width=0.4\linewidth,height=0.33\linewidth]{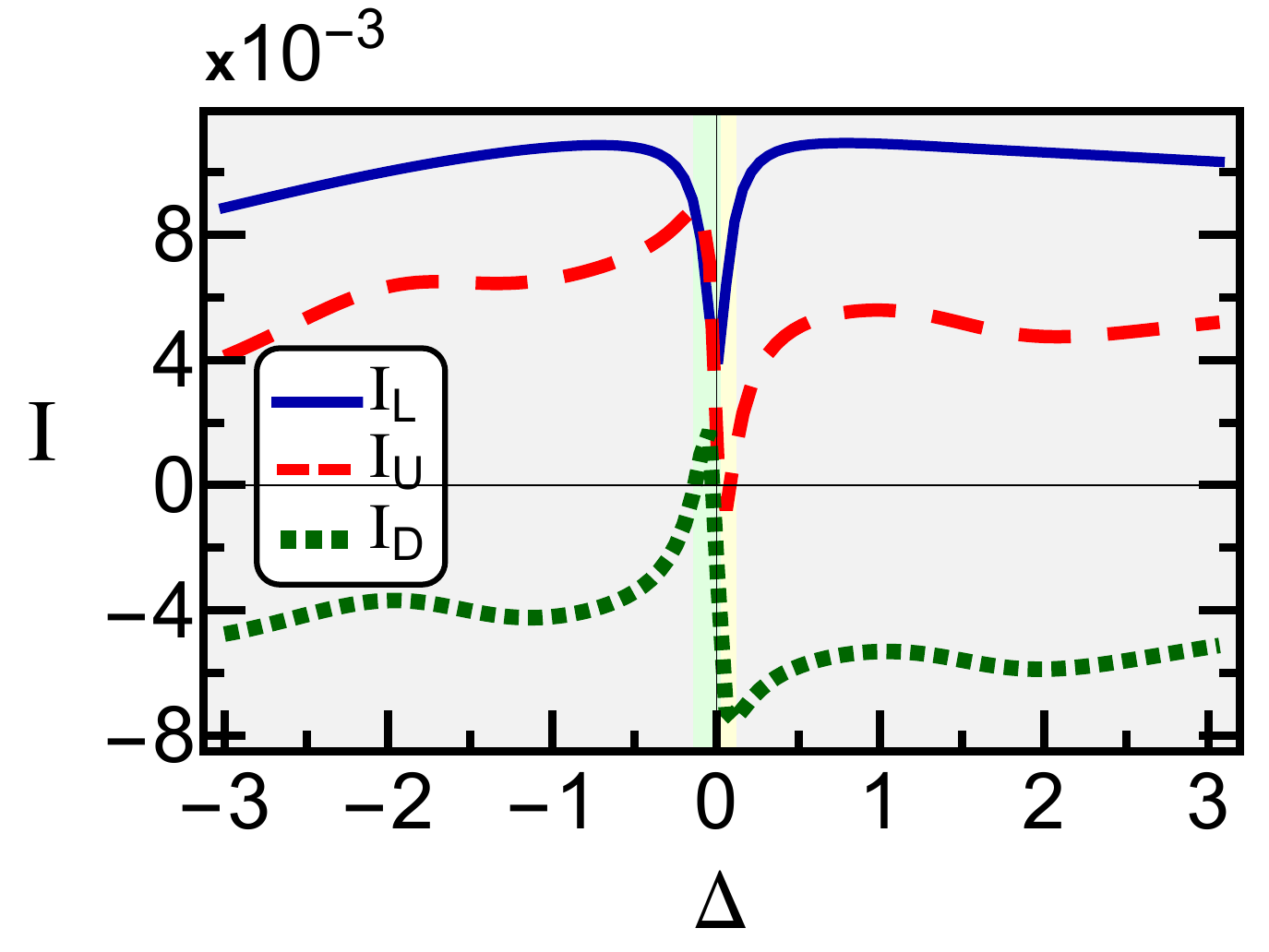}}
		\subfigure[]
		{\includegraphics[width=0.4\linewidth,height=0.36\linewidth]{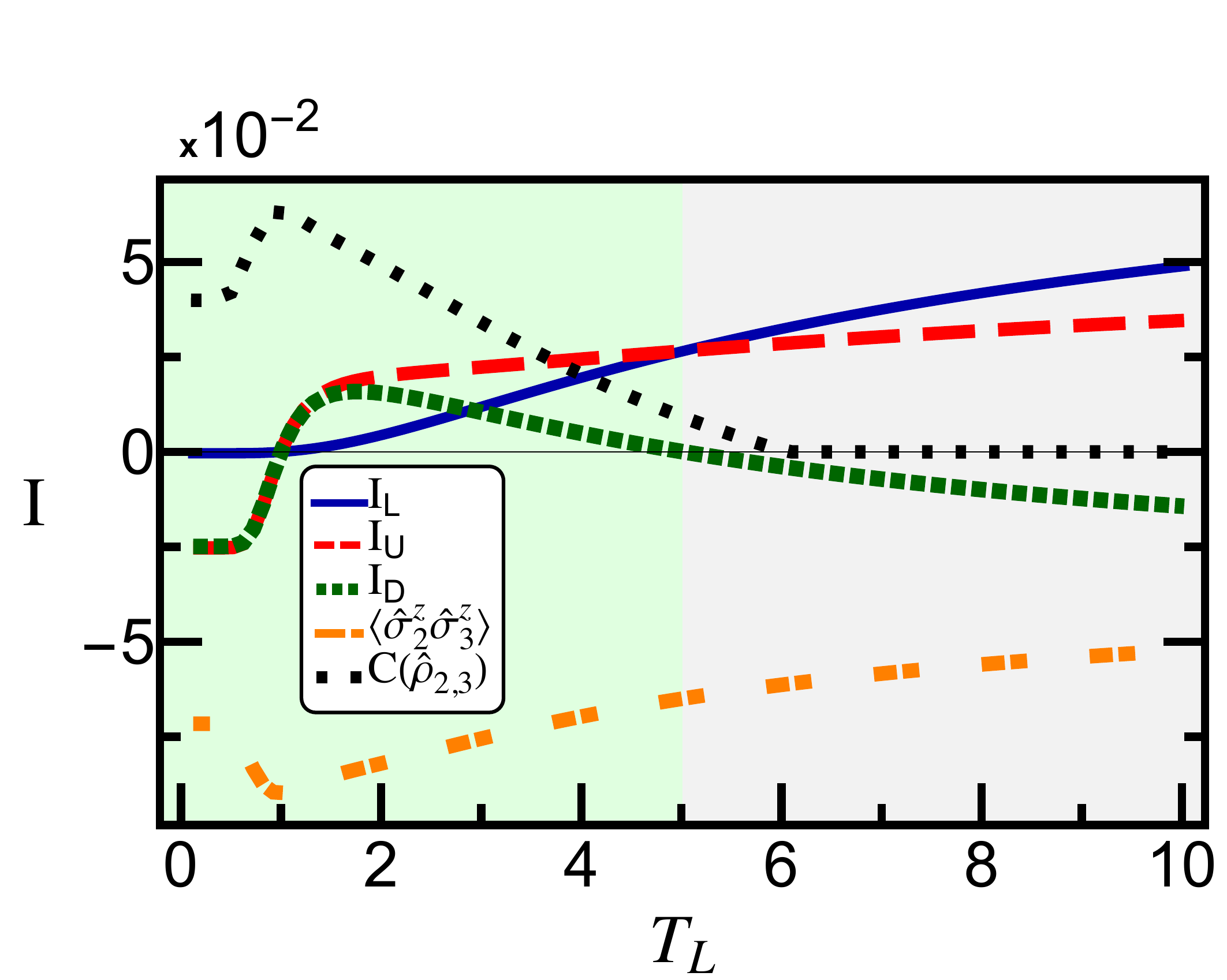}}
		\caption{(Color Online) Variation of Heat Currents ($I$) with the asymmetry parameter $\Delta$ for \textbf{(a)} for setup 5(3$\uparrow$$2\downarrow$)  \textbf{(b)} for setup  5($2\uparrow3\downarrow$), \textbf{(c)} for a 6 spin system with setup   6(4$\uparrow$$2\downarrow$) with $N_U=3, N_D=1$ and \textbf{(d)} Variation of Heat Currents ($I$), correlation and  concurrence between spins 2 and 3  with $T_L$ for $ \Delta=2.1$ . For all the cases unless otherwise specified we work with the setup $5(3\uparrow2\downarrow)$  and  $T_L=2$, $T_R=1$, $J=1$. In \textbf{(d)} we scale the correlation and concurrence by a factor $0.2$ for the visual convenience.}
		\label{Quantum_figures}
	\end{figure*}
	\subsection{Definitions}\label{subsec::Definition_Quantum} 
	{We now define the physical quantities required for studying the Quantum model. Similar to the classical case, all the results are calculated for the steady state and we denote our steady state density matrix as},
	\begin{align} 
		\hat{\rho}_S=\lim_{t\to\infty} \hat{\rho}(t).
	\end{align}
	The expression for the current flowing through a branch is derived by using the Heisenberg equation of motion \cite{breuer2002} for the bond energy operator between spins $l$ and $l+1$.
	The resulting operator for the current flowing through the site `$l$' of our system is given as \cite{main_reference}:
	\begin{align}\label{Quantum_Branch_Current}
		\hat{I}[l]=i[\hat{h}_{l-1,l},\hat{h}_{l,l+1}],
	\end{align}
	with the bond energy operator inferred from the Eq. \eqref{hamiltonian} as,
	\begin{align}
		\hat{h}_{l,l+1}&=J(\hat{\sigma}_x^l \sigma_x^{l+1}+\hat{\sigma}_y^l \hat{\sigma}_y^{l+1})+\Delta\hat{\sigma}_{z'}^l \hat{\sigma}_{z'}^{l+1}.
	\end{align}
	{  Thus, the average steady state branch heat currents become:
		\begin{align}
			I_{U(D)}=Tr(\hat{I}[l]\cdot\hat{\rho}_S),
		\end{align}
		where  $I_{U(D)}$ is the heat current in the upper (lower) branch depending on the position of the $l^{th}$ spin.
		Average current flowing out of a bath is derived from the Liouville operator and is given as \cite{diode6,my}:
		\begin{equation}\label{Quantum_Total_Current}
			\langle I_L\rangle=Tr[\mathcal{L}_{L}\hat{\rho}_S \hat{H}_S].
		\end{equation} 
		Magnetisation of the system is defined with respect to the $\hat{\sigma }_z$ operator of each spin with the excited eigenstate having magnetisation `+1' and the ground eigenstate having magnetisation `-1'. Total magnetisation is the difference between the  spins in excited  state and spins in ground state. To avoid confusion we will use the notation $N_s(N_E \uparrow, N_G \downarrow)$ where $N_s$ is the total number of spins, $N_E (N_G)$ is the number of spins in excited (ground) eigenstate. The average magnetisation of a spin `i' is defined as:
		\begin{align}
			M_i=Tr(\hat{\sigma}_i^z\cdot\hat{\rho}_S).
		\end{align}
		The spin-spin correlation along $`z'$ direction is given as:
		\begin{align}
			\langle \hat{\sigma}^z_i\hat{\sigma}^z_j \rangle=Tr(\hat{\sigma}_i^z\hat{\sigma}^z_j\cdot\hat{\rho}_S).
		\end{align}
		We also use the concurrence \cite{Wooters} as a measure of entanglement between  two spins, it is given by the following formula,
		\begin{align}\label{con}
			C(\hat{\rho})=\text{max}[0,\lambda1-\lambda2-\lambda3-\lambda4],
		\end{align}
		where $\lambda$'s are the eigenvalues in decreasing order of the matrix $\hat{R}=\sqrt{\sqrt{\hat{\rho}} \tilde{\rho}\sqrt{\hat{\rho}}}$
		and  $\hat{\rho}$ is the reduced density matrix for the two spin subsystem of interest with $\tilde{\rho}=(\hat{\sigma}_y \otimes \hat{\sigma}_y) \hat{\rho}^*(\hat{\sigma}_y \otimes \hat{\sigma}_y)$ and $\hat{\sigma}_y$ being  the usual `Pauli Y' matrix. We will discuss in detail how the current magnification influences the correlations between the spins and the concurrence later.
		
		Finally, we  introduce the concept of  ergotropy \cite{ergotropy_defined}, which is defined as the maximum amount of work extractable from a quantum state under unitary time evolution.  { The state which is attainable by the unitary transformation and through which no work is extractable is called the passive state (for example an equilibrium state). To find it, we first need to write the spectral decomposition of the steady state density matrix as well as the system Hamiltonian in the following manner:
			\begin{align}
				\hat{\rho}_S&=\sum_{i} r_i |\rho_i\rangle \langle \rho_i| ,&& \text{with } r_{i}\geq r_{i+1}\nonumber \\
				\hat{H}_s&=\sum_i E_i |E_i\rangle \langle E_i|,&& \text{with } E_{i}\leq E_{i+1}
			\end{align}
			where `$r_i , E_i$' and `$|\rho_i\rangle, |E_i\rangle$' are the eigenvalues and corresponding eigenvectors of the density matrix and the system Hamiltonian respectively.
			Given the above decompositions, the passive state $(\hat{\rho}_{passive})$ is defined as \cite{main_reference,ergotropy_defined},
			\begin{align}\label{passive_state}
				\hat{\rho}_{passive}=\sum_{i}r_i |E_i\rangle \langle E_i|.
			\end{align}
			As can be seen from the above expressions, passive state is not uniquely defined for degeneracies in energy spectrum of the system due to permutations of degenerate energy states.	
			However, the ergotropy $(\epsilon)$ can still be uniquely defined and is given by,
			\begin{align}\label{ergo}
				\epsilon&=Tr(\hat{H}_S\cdot\hat{\rho}_S)-Tr(\hat{H}_S\cdot\hat{\rho}_{passive})\nonumber\\
				&=\sum_{i,j} r_{i}E_j\big{(}|\langle  \rho_i |E_j \rangle|^2-\delta_{i,j}\big{).}
			\end{align}
			From the above discussion, we can infer that ergotropy can be used to quantify how far a system is from the passive state.}
		\begin{figure*}[t]
			\subfigure []
			{\includegraphics[width=0.4\linewidth,height=0.33\linewidth]{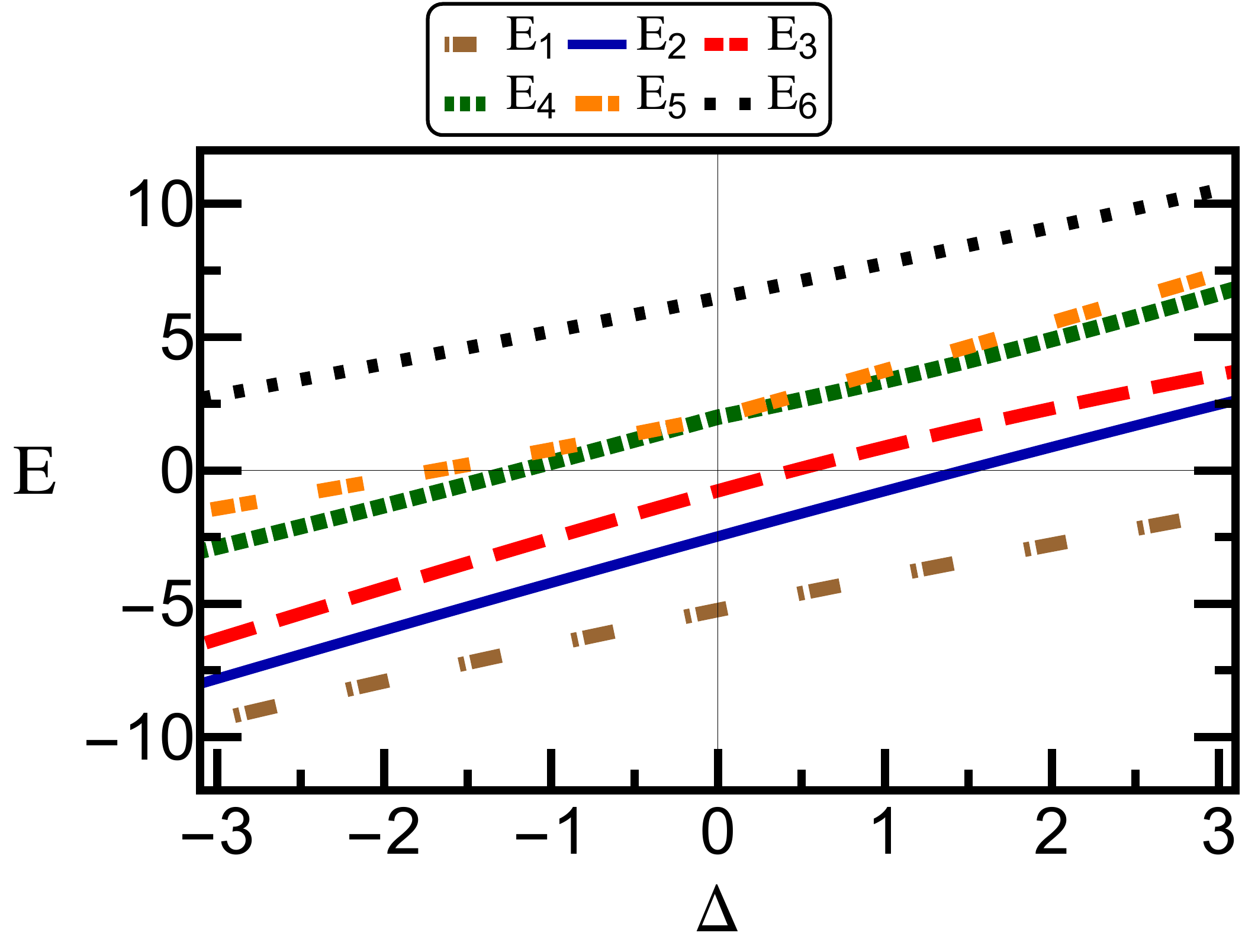}}
			\subfigure []
			{\includegraphics[width=0.4\linewidth,height=0.315\linewidth]{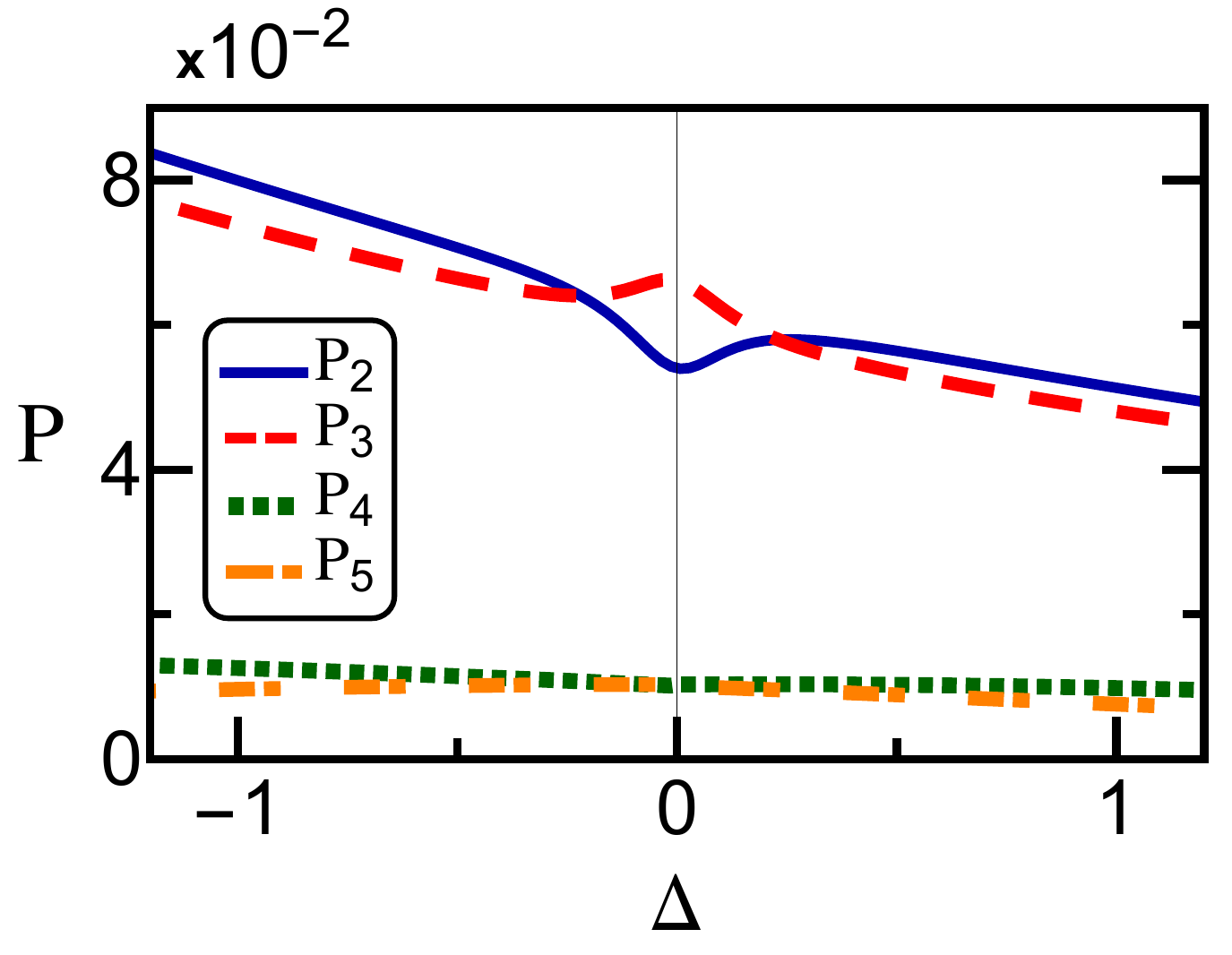}}
			\subfigure[]
			{\includegraphics[width=0.4\linewidth,height=0.33\linewidth]{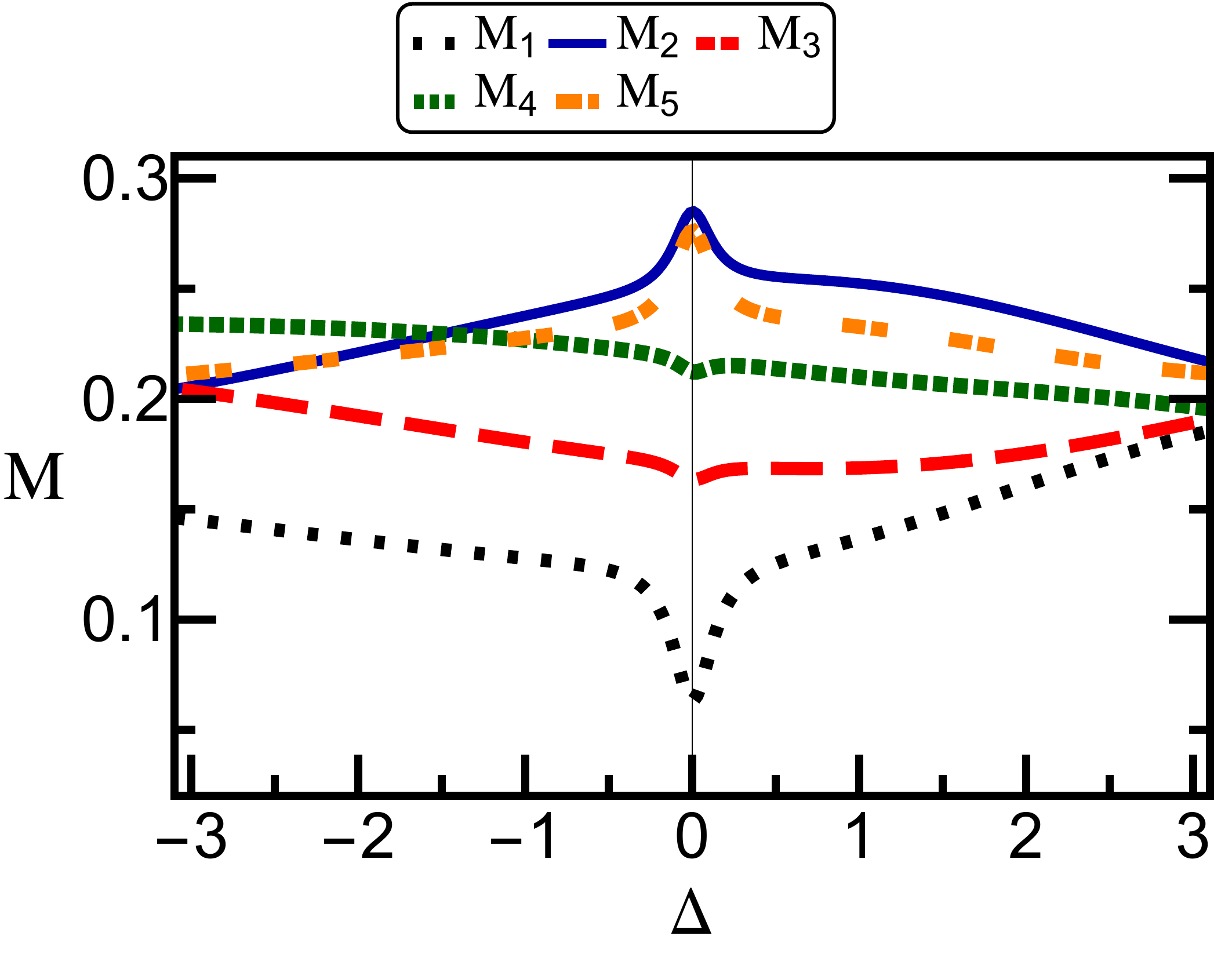}}
			\subfigure[]
			{\includegraphics[width=0.4\linewidth,height=0.315\linewidth]{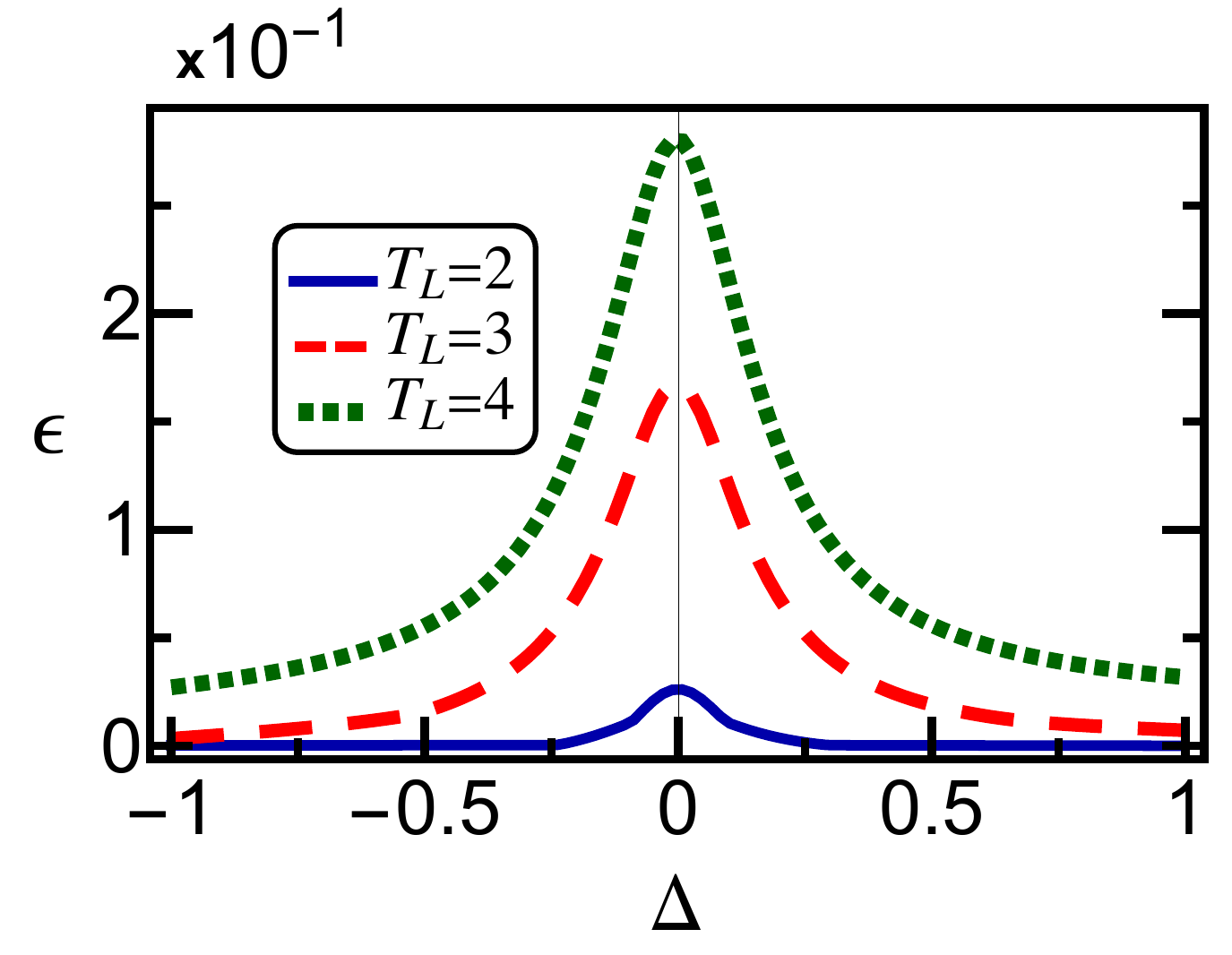}}			
			\caption{(Color Online) \textbf{(a)} Variation of energy levels with $\Delta$, \textbf{(b)} Variation of  occupation probability ($P$) with $\Delta$, \textbf{(c)} Variation of individual spin magnetisation ($M_k$) with $\Delta$ and \textbf{(d)}  Variation of ergotropy ($\epsilon$) with $\Delta$. For all the cases unless otherwise specified, we work with the setup $5(3\uparrow2\downarrow)$ and  $T_L=2$, $T_R=1$, $J=1$. }
			\label{Quantum_figures_explain}
		\end{figure*}
		
		{
			\subsection{Information about numerical methodology}\label{subsec::Definition } 
			For solving the master equation given in Eq. \eqref{Redfield},
			we again write it in the matrix form, though we use the more effective numerical method for the Redfield master equation defined in  \cite{faster_redfield}. Numerically solving this equation by replacing the time derivative terms with $`0'$ and imposing the constraint of fixed initial magnetisation then gives us the steady state density matrix \cite{methods}. Once we have the steady state density matrix, all the quantities of interest can be evaluated from the definitions given in the subsection \ref{subsec::Definition_Quantum}.
			It is important to note that the Redifield master equation  is not guaranteed to be trace preserving and totally positive \cite{breuer2002}, so we also check if the obtained solutions have negative probabilities.  For the parameter range we work on, no such regions  were found. We now discuss in detail the results for the Quantum model below.} 
		\subsection{Results}\label{subsec:: Quantum Results}
		{ 
			The results for the quantum model are shown in Fig. \ref{Quantum_figures}. We see that  large current magnification can be generated just by having unequal number of spins in the branches if the total magnetisation is conserved and the asymmetry parameter `$\Delta$' is in the suitable range. In Fig \ref{Quantum_figures}(a), on plotting the currents with the asymmetry parameter $\Delta$ for the $5$ spin setup $5$($3\uparrow2\downarrow$),  we see that  current magnification occurs for positive values of asymmetric interaction parameter $`\Delta'$. We also see that the onset of current magnification is marked by a sudden dip in the total current near $\Delta\sim0.$ Interestingly on reversing the magnetisation via a configuration change {  $5(3\uparrow 2\downarrow)\to 5(2\uparrow 3\downarrow)$}, the total current remains the same and current magnification also occurs for a certain parameter range but the branch currents drastically change as seen in Fig. \ref{Quantum_figures}(b).
			\begin{figure*}[t]
				\subfigure []
				{\includegraphics[width=0.4\linewidth,height=0.365\linewidth]{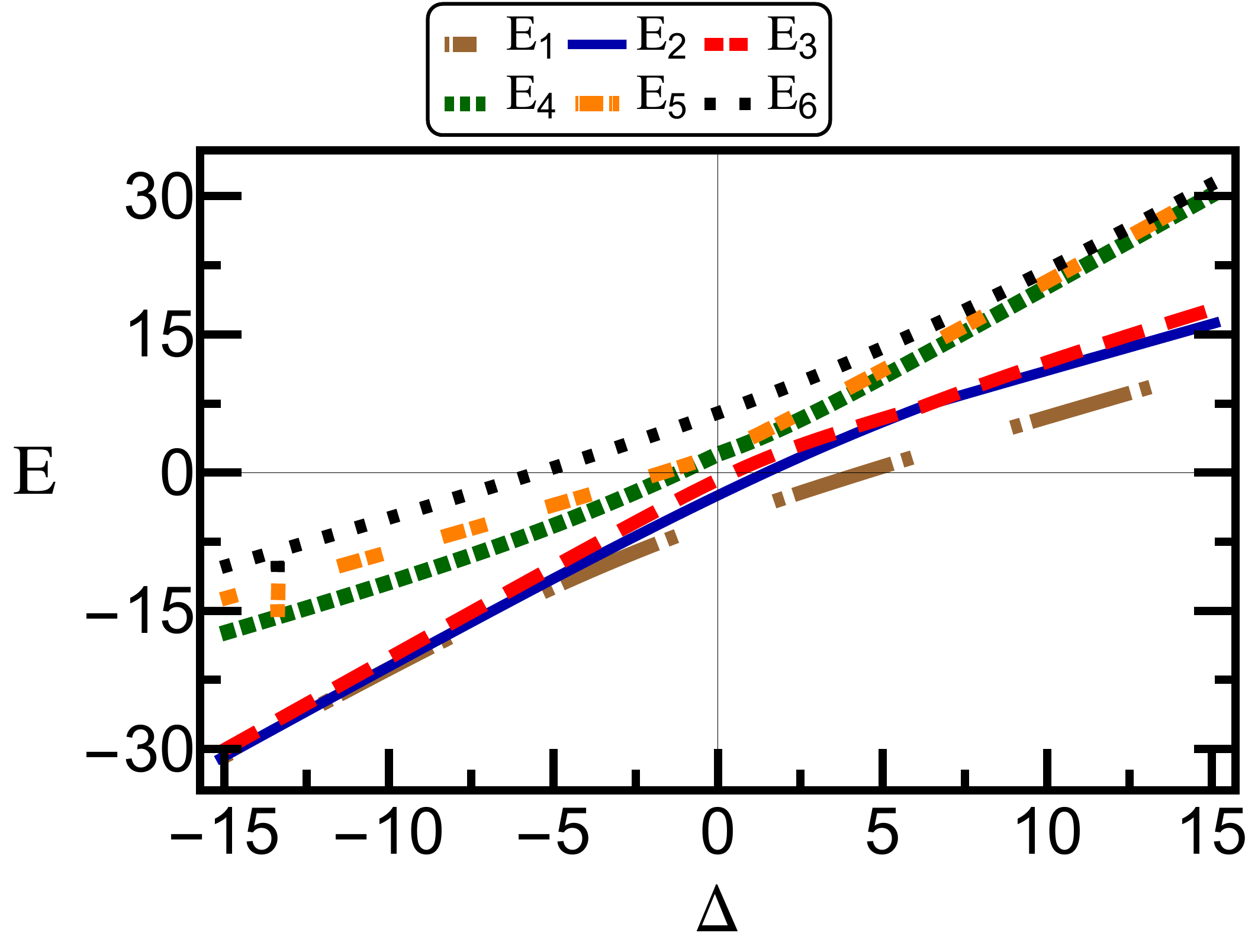}}
				\subfigure []
				{\includegraphics[width=0.4\linewidth,height=0.33\linewidth]{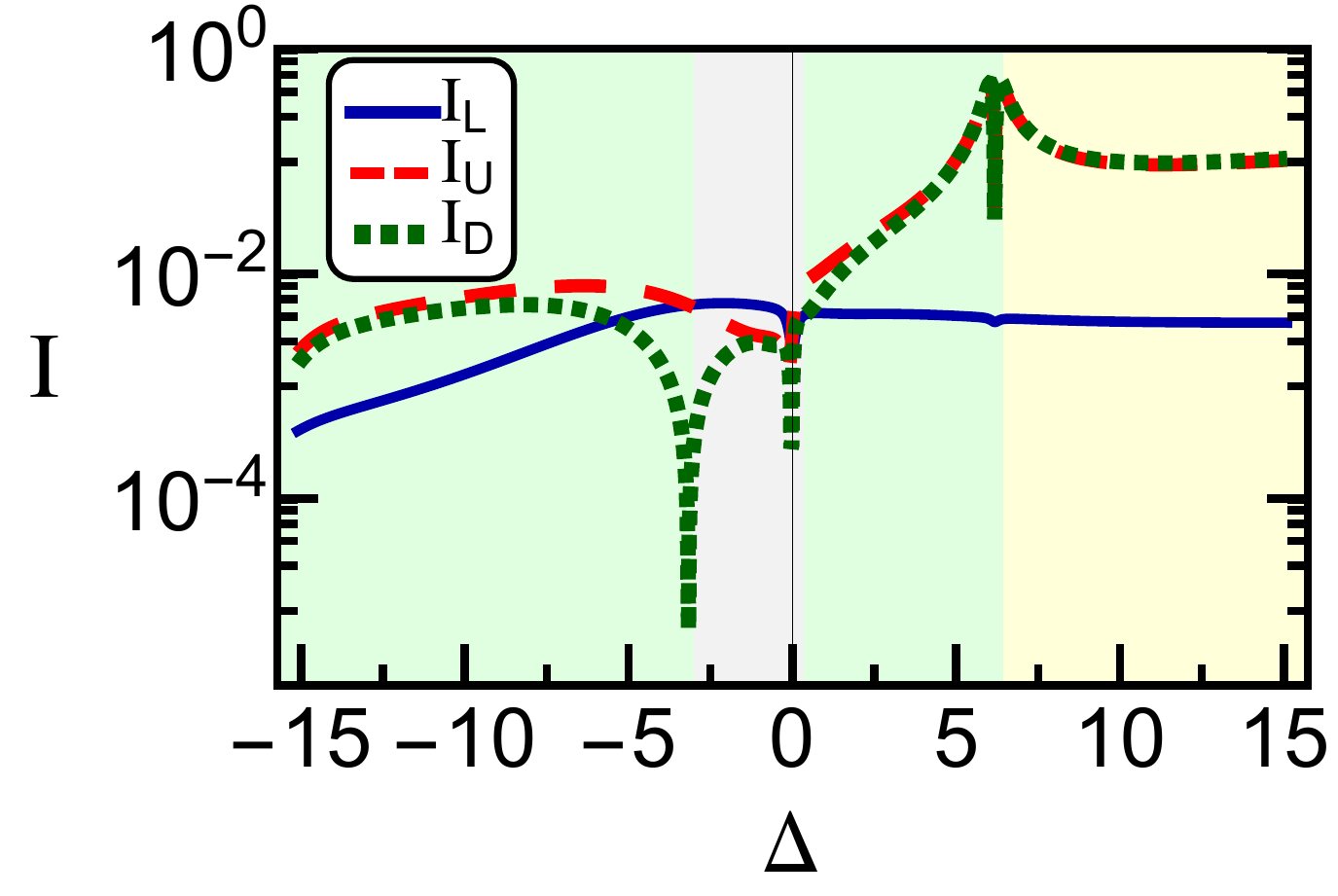}}
				\caption{(Color Online) Variation of \textbf{(a)} energy levels and \textbf{(b)} currents with a large range of asymmetry parameter $\Delta$.  Note that only the magnitude of currents are considered for the log plot in \textbf{(b)} but the background colors for parallel, clockwise or anticlockwise circulating currents are kept. For all the cases unless otherwise specified, we work with the setup $5(3\uparrow2\downarrow)$ and  $T_L=2$, $T_R=1$, $J=1$. }
				\label{Quantum_Range_figures}
			\end{figure*}
			In Fig. \ref{Quantum_figures}(c), we observe that the dip in the total current is sharper for a $6$ spin system and is accompanied by current magnification in a narrow range. Finally on plotting the currents with the temperature of the left bath `$T_L$' while keeping the temperature of right bath `$T_R$' fixed in Fig.  \ref{Quantum_figures}(d), we see that the magnitude of currents initially increase non-monotonically with an increase in the value of `$T_L$', but they plateau for larger temperature differences. On the other hand, the corresponding current magnification initially increases for a small region, reaches a maximum and then starts decreasing before finally vanishing at some finite temperature difference. We also look at the spin-spin correlation along `$z$' direction  and concurrence between spins 2 and 3 in the Fig. \ref{Quantum_figures}(d) to check the dependence of current magnification on correlations. We see that similar to the current magnification, correlation and concurrence initially increase for a small region and then start declining with an increase in temperature of the left bath. Note that $\langle \hat{\sigma}^z_2\hat{\sigma}^z_3\rangle$ is negative because of the antiferromagnetic interaction between the spins. {We try to make sense of the above observations in the following discussion.}
			\par{ We discuss only about the $5$ spins setup $5$(3$\uparrow$2$\downarrow$). The results in Fig. \ref{Quantum_figures} above showed that the onset of current circulation is accompanied by a dip in total current near $\Delta \sim0$. To understand why the total current drops, we study the properties of the occupied energy levels. It turns out that the non-zero occupancy is seen in only $10$ energy levels. These levels correspond to $10$ different configurations having the same magnetisation. Out of these, only $6$ energy levels are non-degenerate, numbering these energy levels in ascending order of their energies we plot $E_1, E_2, E_3, E_4, E_5, E_6$ with $\Delta$ in Fig. \ref{Quantum_figures_explain}(a) and observe that the energy levels $E_4$ and $E_5$ intersect each other at $\Delta =0$. The other energy levels are sufficiently far from each other in the plotted region. This means that the system looses one phonon transfer channel through which heat current can pass at $\Delta=0$ and as a result the total heat current decreases. To check what happens to the  probability of occupancy corresponding to these energy levels, we plot the variation of probability of occupancy with $\Delta$ for the same range in Fig. \ref{Quantum_figures_explain}(b). We see that  $P_2$ suddenly drops and $P_3$ increases indicating population inversion between energy levels $E_2$ and $E_3$ near $\Delta=0$. The probability of occupancy for other energy levels behaves in the expected way with higher energy levels having lesser probability of occupancy and vice versa. This can only happen if the system is far from the passive state defined in \eqref{passive_state}  and seems to suggest that the additional degeneracy in the two energy levels and the constraint of fixed initial magnetisation is forcing the system to settle at states that are unfavorable according to a passive state. We also plot the variation of individual spin magnetisation with $\Delta$ in Fig. \ref{Quantum_figures_explain}(c) and see that though the total magnetisation remains fixed throughout this region, individual spin magnetisation behave in an atypical manner near $\Delta \sim 0$ with local minima or maxima dipping for each spin. To support our observations further, we evaluate the ergotropy according to the Eq. \eqref{ergo} but only consider the energy levels with non zero probability of occupancy. Plotting the variation of ergotropy with $\Delta$ in Fig. \ref{Quantum_figures_explain}(d), we see that the ergotropy suddenly starts rising around $\Delta\sim0$ indicating the system does move farther away from the passive state in this region. Also, on increasing the temperature of the left bath, the ergotropy increases as the system moves even further away from the passive state due to the increase in the temperature gradient. }
			
			\par All the above results seem to suggest that whenever two energy levels intersect, the physical observables of our system behave in an atypical manner. To further probe this, we study the variation of energy levels and currents for a larger range of the asymmetric parameter $\Delta$ in Fig. \ref{Quantum_Range_figures}. We see in Fig. \ref{Quantum_Range_figures}(a) that only in the region where $\Delta$ is negative with small magnitude $(-2.5<\Delta<0)$, all the energy levels have distinct values. In Fig. \ref{Quantum_Range_figures}(b), we see that this region also corresponds to the only parameter range where we get parallel currents in the branches. Moving on either side of this region leads to the energy levels coming closer to each other and the transition from parallel to circulating currents. Specifically in Fig. \ref{Quantum_Range_figures}(a) for even more negative values of $\Delta(\sim-10)$, the energy levels $E_2, E_3$ and $E_1$ move towards each other. On the other side, as we move to positive values of $\Delta(\sim5)$, initially $E_4$, $E_5$ move towards each other followed by $E_6$ coming closer to them for even larger positive values of $\Delta(\sim11)$. Correspondingly in \ref{Quantum_Range_figures}(b), we see that there is a transition from parallel current to clockwise circulating current as we move away from $\Delta=0$ on either side, though the regions are not symmetric with respect to $\Delta$. Also, on moving more towards the positive side of  $\Delta(\sim6)$, there is a point of transition from clockwise circulating current to anticlockwise circulating current. This point closely corresponds to the intersection of energy levels $E_2$ and $E_3$. Considering the above observations, it appears that whenever two energy levels come towards each other,  it can lead to a transition from parallel to circulating currents in our system.
			
			\par However, it is important to note that the current magnification discussed above is only possible for temperature gradients of the same order as the interaction strengths. In the high temperature regime current magnification is absent as the inter-spin correlations are small and the currents just flow parallel in the branches (see Fig. \ref{Quantum_figures}(d)). Though the above discussion does help us to predict suitable conditions for current circulation namely ergodic constraints and additional degeneracies. We note that analysing the phonon transfer mechanism between different energy levels might help us to get better insights into current magnification in quantum systems but this is beyond the scope of the present manuscript \cite{my,diode6}.}}
	\section{Conclusion}\label{sec::Conclusion}
	To conclude, we study heat current magnification due to branch spin number asymmetry in  classical and quantum spin systems. We find that current magnification is absent in the classical Q2R model for symmetric branch interaction strengths. However, we encounter an interesting feature that the heat current { flowing through a spin branch is inversely proportional to the number of spin-spin bonds in that branch}. We employ different spin-spin interaction strengths in the upper and lower branch and show that this inequality is enough to generate as well as manipulate the current magnification. This happens even if we have the same number of spins in both the branches. We then provide a possible physical mechanism responsible for current magnification in such systems. Similar features and { heat current values are found if we use the CCA model instead of the Q2R model. This shows that the presence of additional momentum energy in the system does not contribute appreciably to the steady state currents. This is in accordance with the earlier studies with such dynamics}. We then study a five spin Quantum system with modified Heisenberg XXZ interaction and preserved magnetisation using the Redfield master equation. We, with detailed numerical analysis, show that it is possible to generate current magnification just by the branch spin number asymmetry in this model for a suitable range of asymmetry parameter $\Delta$. Our results indicate that the onset of current magnification is accompanied by a sudden dip in total current which may be triggered due to the { intersection of two energy levels for certain values of $\Delta$. It is seen that this value of $\Delta$ also corresponds to population inversion and atypical trends for other physical observables e.g. magnetisation.
		Since the two factors namely the ergodic constraint due to fixed magnetisation and additional degeneracies in the system push the system away from the passive state, we deduce that they may be the main causes for current magnification. This deduction is also supported by the ergotropy evaluations for our system.} It is also seen that the total current is immune to the inversion of individual magnetisation of the spins while the branch currents are not. In both the classical as well as quantum case, a point exists where the current in one of the branch is zero and the total current flows only through one of the branches. This point corresponds to the transition between parallel and circular currents. {We also find that for both the classical and quantum models, current magnification is only observed when temperature gradient and intra-system interaction strength have similar order of energy. This also points to the importance of correlation between spins required for current magnification.}  In further studies, the relation of current magnification with ergodicity can be studied as both the classical and quantum systems considered here have non ergodic dynamics \cite{Q2R_STAUFFER2000113, CREUTZ198662}, the effect of different spectral densities can also be explored in the quantum case.
	\section{acknowledgement}
	RM gratefully acknowledges financial support from Science and Engineering Research Board (SERB), India, under the Core Research Grant (Project No. CRG/2020/000620). We also thank \"Ozg\"ur E. M\"ustecapl\ifmmode \imath \else \i \fi{}o\ifmmode \breve{g}\else \u{g}\fi{}lu and M. Tahir Naseem of the  Ko\c{c} University, Istanbul for useful discussions and insights in this project.

	\bibliography{circulation}
	
\end{document}